\documentclass[article]{jss}

\PassOptionsToPackage{usenames,dvipsnames}{xcolor}

\usepackage{amsfonts}
\usepackage{amssymb}
\usepackage{amsthm}
\usepackage{amsmath}
\usepackage{xcolor}
\usepackage{bm}
\usepackage{ulem}
\usepackage{tikz}
\usetikzlibrary{trees}
\usepackage{framed}
\usepackage{hyperref}
\usepackage[utf8]{inputenc}
 
\newcommand{\M}{\mathfrak{m}}

\author{Florian Frommlet\\Medical University \\ of Vienna
        \And 
        Jon Lachmann\\Stockholm University \And 
        Geir Storvik\\University of Oslo
        \And 
        Aliaksandr Hubin\\NMBU, KBM\\ and \\ University of Oslo}
\title{\pkg{FBMS}: An R Package for Flexible Bayesian Model Selection and Model Averaging}

\Plainauthor{Florian Frommlet, Jon Lachmann, Geir Storvik, Aliaksandr Hubin} 
\Plaintitle{Flexible Bayesian Model Selection FBMS} 
\Shorttitle{\pkg{FBMS}: Flexible Bayesian Model Selection} 

\Abstract{ 
The \pkg{FBMS} R package facilitates Bayesian model selection and model averaging in complex regression settings by employing a variety of Monte Carlo model exploration methods. At its core, the package implements an efficient Mode Jumping Markov Chain Monte Carlo (MJMCMC) algorithm, designed to improve mixing in multi-modal posterior landscapes within Bayesian generalized linear models. In addition, it provides a genetically modified MJMCMC (GMJMCMC) algorithm that introduces nonlinear feature generation, thereby enabling the estimation of Bayesian generalized nonlinear models (BGNLMs). Within this framework, the algorithm maintains and updates populations of transformed features, computes their posterior probabilities, and evaluates the posteriors of models constructed from them.
We demonstrate the effective use of \pkg{FBMS} for both inferential and predictive modeling in Gaussian regression, focusing on different instances of the BGNLM class of models. Furthermore, through a broad set of applications, we illustrate how the methodology can be extended to increasingly complex modeling scenarios, extending to other response distributions and mixed effect models.
} 

\Address{
  Aliaksandr Hubin\\
  KBM\\
  Norwegian University of Life Sciences\\
1432 Ås, Norway\\
  E-mail: \email{aliaksandr.hubin@nmbu.no}\\
  URL: \url{https://www.nmbu.no/en/about/employees/aliaksandr-hubin}
}

\begin{document}

\section[Introduction]{Introduction}

Bayesian model selection and model averaging are essential tools in the analysis of complex regression models, offering a robust framework for handling uncertainty in model structures and improving predictive performance. Traditional approaches often face challenges in exploring the vast model spaces efficiently, particularly in multi-modal settings or when incorporating nonlinearities. The Flexible Bayesian Model Selection (\pkg{FBMS}) R package was developed to address these challenges through innovative Markov Chain Monte Carlo exploration algorithms.

\pkg{FBMS} builds upon the rich class of Bayesian generalized nonlinear models (BGNLMs), leveraging mode-jumping MCMC (MJMCMC) and its genetically modified variant (GMJMCMC). These algorithms enable efficient navigation of complex, multi-modal model spaces, incorporating advanced feature transformations and interactions. The package supports a wide range of model types, including generalized linear and nonlinear models, mixed-effects models, and survival models, making it a versatile tool for statisticians and data scientists.

This paper introduces the \pkg{FBMS} package and situates it within the broader landscape of Bayesian modeling tools. Compared to existing software, \pkg{FBMS} offers unique advantages in its flexibility. The package's design facilitates additional customization, allowing users to incorporate their own priors, likelihoods, random effects, and feature-generation strategies. The effectiveness and adaptability of the \pkg{FBMS} package will be  demonstrated in a wide range of applications, including the recovery of physical laws and developing predictive models. With \pkg{FBMS} it becomes fairly convenient to perform certain modeling tasks which might be quite hard to achieve with other existing software, like for example Bayesian  mixed-effects Poisson regression combined with fractional polynomials in Section \ref{SubSec:MixedPoisson}.

The rest of the manuscript is organized as follows: Section \ref{fbmsec} outlines the theoretical underpinnings of \pkg{FBMS} and its core algorithms. Section \ref{Sec:BasicIntro} makes use of a simple example to introduce the basic aspects of the \pkg{FBMS} package.  Section \ref{Sec:Metric} explores how to specify different subclasses of nonlinear features in the context of the Gaussian model, i.e. linear regression with Gaussian errors. In Section \ref{Sec:Priors}, we show how to specify custom-made priors for the Gaussian model. More general models will be discussed afterwards in Section \ref{Sec:Extensions},  including logistic regression, linear mixed model, mixed effects Poisson regression, and Cox regression for survival data. 
After a brief discussion section some more specific computational details of the algorithm are described in the Appendix. 

The \proglang{R} package for this version of the manuscript can be installed using the following command:

\code{devtools::install_github("jonlachmann/FBMS@v1_arxiv", build_vignettes=F)}.

And all \proglang{R} code for all examples discussed in this manuscript is available as supplementary material at \url{https://github.com/jonlachmann/FBMS/tree/v1_arxiv/tests_current}. Several of these examples, though not all of them are also included in the vignette of the \pkg{FBMS} package. To get the vignette available, please install with \code{build_vignettes=T} option (the installation in this case will be notably more time consuming).

\section[FBM]{Flexible Bayesian Modeling}\label{fbmsec}

The \pkg{FBMS} package is based on the rich class of nonlinear Bayesian models called BGNLM (Bayesian generalized nonlinear models) which was introduced by \citet{hubin2021flexible}. 
For the sake of simplicity we will formulate here the basic ideas assuming Gaussian responses. Extensions to other response distributions will be discussed in Section \ref{Sec:Extensions}, including examples from the exponential family,  models with random effects and survival models. 

\begin{figure}
    \centering
\begin{tikzpicture}[
  every node/.style={align=center, rounded corners, draw=black, fill=white, thick},
  level 1/.style={sibling distance=50mm},
  level 2/.style={sibling distance=30mm},
  level 3/.style={sibling distance=15mm}
]

\node {sin}
  child { node {+}
    child { node {1} }
    child { node {*}
      child { node {2} }
      child { node {$x_1$} }
    }
    child { node {*}
      child { node {3} }
      child { node {exp}
        child { node {+}
          child { node {5} }
          child { node {*}
            child { node {$x_2$} }
            child { node {$x_3$} }
          }
        }
      }
    }
    child { node {+}
      child { node {cos}
        child { node {$x_4$} }
      }
    }
  };

\end{tikzpicture}
\caption{Tree representation of the feature \( \sin(1 + 2*x_1 + 3*\exp(5 + x_2*x_3) + \cos(x_4)) \).}\label{figure:featuretrees}
\end{figure}
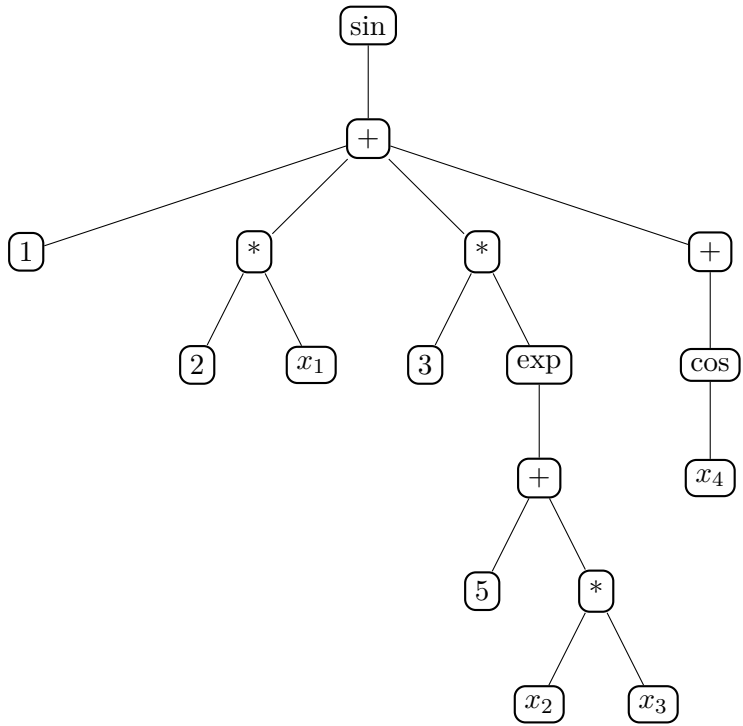

Assume that there are $n$ observations of a metric response variable $Y_i,  i \in \{1,\dots,n\}$ and for each observation there is a $p$-dimensional vector of input covariates $\bm{x}_{i} = (x_{i1}, \dots, x_{ip})$. 
We consider a model with i.i.d. normally distributed error terms $\epsilon_i \sim \mathcal{N}(0, \phi)$, with $\phi = \sigma^2$ being the dispersion parameter, where we allow for a large class of nonlinear expressions (features $F_{j}(\bm{x}_i) \subset \mathcal{F}$) as regressor variables:
\begin{equation}\label{themodeleq}
  Y_i = \beta_0  + \sum_{j=1}^{q} \gamma_{j}\beta_{j}F_{j}(\bm{x}_i) + \epsilon_i \; .
\end{equation}
The class of potential features $\mathcal{F}$ is vast, but the subset of features that enter a specific model is assumed to be fairly moderate. 
Here, $\gamma_j$ is an indicator variable which determines whether the $j$-th feature $F_{j}(\bm{x}_i) \subset \mathcal{F}$ enters the model.
The class $\mathcal{F}$ includes all original covariates $\bm{x}_i$ and is then built up sequentially using nonlinear transformations (operators) to generate new features. Consequently features can be seen as functional trees 
using a predefined set of nonlinear transformations. A tree representation of a specific feature is shown in Figure~\ref{figure:featuretrees}. By restricting the trees to a certain predefined maximal depth, this process induces a finite total number $q$ of potential feature sstructures and thus $2^q$ candidate models. 

Within the \pkg{FBMS} package,
the class $\mathcal{F}$ is constructed through
three different operators. 
A nonlinear transformation can be either applied directly to already existing features ({\it nonlinear modifications}) or, like in neural networks, applied to linear projections of existing features ({\it nonlinear projections}). To this end, the user can specify a set $\mathcal{G}$ of nonlinear functions from which the algorithm randomly selects to generate new features.   Furthermore one can build {\it interactions} of existing features.  The iterative application of these three  nonlinear operations  can be used to generate features from a vast functional space \citep[see][for further details]{hubin2021flexible}. Depending on the choice of nonlinear functions in $\mathcal{G}$  and on certain restrictions of the feature generating operators, one obtains many interesting families of nonlinear models as special cases. Some of them are described in more detail in
Section \ref{Sec:Metric}.

\subsection{Prior specification}\label{SubSec:PriorSpec}

In the Bayesian approach, one has to specify priors for the different models as well as for the parameters within each  model. For the latter, the prior distributions for the regression coefficients \(\bm{\beta}\) included in the model and, if applicable, the dispersion parameter \(\phi\), are specified as follows:  
\begin{align}  
p(\bm{\beta},\phi|\M) = p(\bm{\beta}|\M,\phi) p(\phi|\M). \label{eq:prior.par}  
\end{align}  
Here, \(\bm{\beta}\) represents the regression parameters given model \(\M=(\gamma_1,...,\gamma_q)\). A common choice for these parameters is the family of mixture \(g\)-priors, which exhibit desirable properties for Bayesian variable selection and model averaging~\citep{li2018mixtures}. However, alternative priors, such as Jeffreys prior, are also popular. The choice of priors for \(\bm{\beta}\) and \(\phi\) is often guided by computational considerations, particularly in facilitating efficient marginal likelihood estimation (e.g., by leveraging conjugate priors).  
\subsubsection{Priors for regression coefficients}

One of the simplest options is Jeffreys prior, which is both numerically scalable and robust for most of the \pkg{FBMS} applications. It is defined as  
\begin{align}  
p(\bm{\beta}|\phi,\M) \propto |\mathcal{I}_\M(\bm{\beta}, \phi)|^{1/2},  
\end{align}  
where \(\mathcal{I}_\M(\bm{\beta}, \phi)\) is the Fisher information matrix for parameters of model $\M$ with $\phi$ given. This prior is noninformative and is often used in scenarios where objective Bayesian inference is desired.

While Jeffreys prior provides a noninformative specification for \(\bm{\beta}\), it is improper. A widely used alternative 
is the \( g \)-prior, originally introduced by Zellner \citep{zellner1996models}, which assumes a normal distribution for \(\bm{\beta}\),  
\begin{align}  
p(\bm{\beta} | \phi, g, \M) = \mathcal{MVN} \left( 0, g \phi (X_\M^\top X_\M)^{-1} \right) 
\end{align} 
and ensures properness by introducing a scaling factor \( g \). Here $X_\M$ is the design matrix which corresponds to the features specified by $\M$. The \( g \)-prior maintains a covariance structure proportional to the Fisher information matrix, similar to Jeffreys prior,  but regularizes the posterior distribution. By tuning \( g \), one can control the amount of shrinkage imposed on \(\bm{\beta}\), making this prior particularly useful for Bayesian variable selection and model averaging.

 When choosing $p(\phi) \propto \phi^{-1}$ it is well known that for the Gaussian model, up to a constant, the logarithm of the marginal posterior probability  becomes 
\begin{equation}\label{mloglik.gprior}
 \log p(Y|\M) = \mbox{const} + \frac{n - k}2 \log(1+g) -  \frac{n - 1}2 \log(1 + g (1 - R_\M^2)) \; ,
\end{equation}
where $R_\M^2$ is the coefficient of determination corresponding to the estimated model including features specified by $\M$.
For our default choice, we follow~\citet{fernandez2001benchmark} and use $g=\max\{n,p^2\}$.

Furthermore, to mitigate the sensitivity of inference to a fixed \(g\), robust alternatives introduce a hyperprior on \(g\).
This hierarchical structure increases flexibility and reduces the influence of any single fixed value for \(g\).
A common approach is to assume that \(u=1/(1+g)\) follows a truncated compound hypergeometric (tCCH) prior,
\begin{align}
p\left(u\right) &= tCCH\left(\frac{a}{2},\frac{b}{2},\rho,\frac{s}{2},v,k\right)
\end{align}
This family of mixtures of \( g \)-priors covers a wide range of priors examined in the literature; see \citet{li2018mixtures} for a detailed review.
A particularly recommended variant is the robust \( g \)-prior, whose parameters are listed in Table~\ref{tab:priors} along with all other parameter priors available in the \pkg{FBMS} package.
For a broader overview of parameter priors, we refer the reader to \citet{li2018mixtures} and \citet{bayarri2012criteria}.

\begin{table}[!tb]
    \centering
    \small
    \renewcommand{\arraystretch}{1.3}
    \caption{Priors for regression parameters available in the \pkg{FBMS} package, along with their hyperparameters and applicability.
\( p_{\M} \) denotes the number of predictors, excluding the intercept.
“G” indicates applicability to the “gaussian” family only, while “GLM” additionally includes the “binomial”, “poisson”, and “gamma” families.
The hyperparameters \( a, b, \rho, s, v, k \) are defined for the tCCH priors; other priors use their own specific parameters, as noted.
Parameters in italics (e.g., \( a, s, g,\)) denote values specified by the user. Var for Jeffreys-BIC is either set to "unknown" or a fixed positive real number.}
    \label{tab:priors}
    \begin{tabular}{lccccccc}
    \\
        \textbf{Prior (Alias)} & &  &  &  &  &  & \textbf{Families} \\
        \hline
        \multicolumn{8}{l}{\underline{Default:}} \\
        \texttt{g-prior} & \multicolumn{6}{c}{\( \text{g} \) (default: \( \max(n, p^2) \))} &  GLM \\
        \\
        \multicolumn{8}{l}{\underline{tCCH-Related Priors:}} \\
   
       \textit{Parameters} & \( a \) & \( b \) & \( \rho \) & \( s \) & \( v \) & \( k \) &  \\
         \cline{1-7}
         \texttt{CH} & \( a \) & \( b \) & 0 & \( s \) & 1 & 1 &  GLM \\
        \texttt{uniform} & 2 & 2 & 0 & 0 & 1 & 1 &  GLM \\
        \texttt{Jeffreys} & 0 & 2 & 0 & 0 & 1 & 1 &  GLM \\
        \texttt{beta.prime} & \( \frac{1}{2} \) & \( n - p_{\M} - 1.5 \) & 0 & 0 & 1 & 1 &  GLM \\
        \texttt{benchmark} & 0.02 & \( 0.02 \max(n, p^2) \) & 0 & 0 & 1 & 1 &  GLM \\
        \texttt{TG} & \( 2a \) & 2 & 0 & \( 2s \) & 1 & 1 &  GLM \\
        \texttt{ZS-adapted} & 1 & 2 & 0 & \( n + 3 \) & 1 & 1 &  GLM \\
        \texttt{robust} & 1 & 2 & 1.5 & 0 & \( \frac{n+1}{p_{\M} + 1} \) & 1 &  GLM \\
        \texttt{hyper-g-n} & 1 & 2 & 1.5 & 0 & 1 & \( \frac{1}{n} \) &  GLM \\
        \texttt{intrinsic} & 1 & 1 & 1 & 0 & \( \frac{n + p_{\M} + 1}{p_{\M} + 1} \) & \( \frac{n + p_{\M} + 1}{n} \) &  GLM \\
        \texttt{tCCH} & \( a \) & \( b \) & \( \rho \) & \( s \) & \( v \) & \( k \) &  GLM \\
        \\
        \multicolumn{8}{l}{\underline{Other Priors:}} \\
         \texttt{EB-local} & \multicolumn{6}{c}{\( a \)} &  GLM \\
        \texttt{EB-global} & \multicolumn{6}{c}{\( a \)} & G \\
        \texttt{JZS} & \multicolumn{6}{c}{\( a \)} & G \\
        \texttt{ZS-null} & \multicolumn{6}{c}{\( a \)} & G \\
        \texttt{ZS-full} & \multicolumn{6}{c}{\( a \)} & G \\
         \texttt{hyper-g}& \multicolumn{6}{c}{\( a \)} & GLM \\
        \texttt{hyper-g-laplace} & \multicolumn{6}{c}{\( a \)} & G \\
        \texttt{AIC} & \multicolumn{6}{c}{None} & GLM \\
        \texttt{BIC} & \multicolumn{6}{c}{None} &  GLM \\
        \texttt{Jeffreys-BIC} & \multicolumn{6}{c}{Var ("unknown" or numerical value)} &  GLM \\
        \hline
    \end{tabular}
\end{table}

As previously mentioned, our default prior is the $g$-prior with $g = \max(n,p^2)$.
However, in Section~\ref{Sec:Priors} we demonstrate how to use other parameter priors listed in Table~\ref{tab:priors}, as well as how to specify custom priors not implemented in the \pkg{FBMS} package.

\subsubsection{Priors for dispersion parameters}

For dispersion parameters, we employ relatively simple prior specifications.
In the binomial family (as in logistic regression), as well as in the Poisson and gamma families, $\phi$ is fixed at 1.
For the Gaussian distribution, the variance may either be specified as fixed and known or treated as unknown.
When the variance is unknown, the only implemented prior for $\phi = \sigma^2$ within the tCCH class and for the Jeffreys prior is
\begin{equation}
p(\sigma^2)\propto\frac{1}{\sigma^2}.    
\end{equation}
Another option available in \pkg{FBMS} is the empirical Bayes approach as implemented in the \pkg{BAS} package \citep{Clyde:Ghosh:Littman:2010}.

\subsubsection{Model priors}

Next we have to specify priors for the model topology. The features included in a model define its structure, which can be formalized by considering the vector $\M = (\gamma_1,\dots,\gamma_q)$. The specific choice of a prior for the indicator variables $\gamma_j$ is crucial in as much as it should allow for nonlinear modeling but at the same time avoid too much over-fitting to the data. This can be achieved by penalizing  features according to their degree of complexity. 

The default prior of \pkg{FBMS} assumes independence of inclusions of different effects and has the following form:
 \begin{equation}
 p(\M) \propto \prod_j\  r^{\gamma_jc(F_j(\bm x))}\label{glmgammaprior}     
 \end{equation}
with $0<r<1$ and ${c(F_j(\bm x))}\ge 0$ being some complexity measure. Per default we use $r = 1/n$ and
the default choice for $c(F_j(\bm x))$ is the operation count (\code{oc}), which was comprehensively described by \citet{hubin2021flexible}. It essentially reflects the number of algebraic operations required to compute a given feature.
The \pkg{FBMS} package allows users to implement alternative model priors. Section~\ref{SubSec:model_prior} illustrates how to implement different priors based on alternative complexity measures, using a specific example from Bayesian logic regression \citep{ruczinski2003logic}.

\subsection{Posterior probabilites}

Having fully specified the model and its priors, our primary interest lies in posterior probabilities for the model structure $\M$, which is given by 
\begin{equation}\label{modelposterior}
p(\M|Y) =\frac{ p(Y|\M)\ p(\M)}{ \sum_{\M'\in\mathcal{M}}p(Y|\M')\ p(\M')},
\end{equation}
where $\mathcal{M}$ is the set of all possible model configurations.
Estimation of these posterior probabilities 
is performed by a tailored MCMC algorithm, which can handle multiple modes and can easily be parallelized, see Section~\ref{sec:alg} for details. 

Of central importance in \eqref{modelposterior} is the computation of the marginal likelihood $p(Y|\M)$ and its  logarithm \( \log p(Y | \M) \), which will depend on the chosen parameter prior. Here we will provide details about some important special cases. A more extensive treatment can be found in the comprehensive overview given by  \citet{li2018mixtures}.

For Jeffreys prior, the marginal likelihood is computed using Laplace approximation leading essentially to the BIC criterion as log marginal likelihood: 
\begin{align}
    \log p(Y|\M) \approx \mbox{const} + \log p(Y|\hat{\bm\beta},\phi,\M) - 0.5\ p_\M\log n,
\end{align}
Here $p_\M$ is the number of predictors in the model $\M$ (not including the intercept),   $\hat{\bm\beta}$ is the maximum likelihood estimate, and $\phi$ is either estimated at the mode of the likelihood or assumed to be known.  
In case of the standard \(g\)-prior for linear models with Gaussian noise, the integral over \(\bm{\beta}\) can be solved analytically due to the conjugate normal structure, resulting in a closed-form log marginal likelihood of form \eqref{mloglik.gprior}.

Using a tCCH hyperprior for $g$, precise integrated Laplace approximations of the marginal likelihood for generalized linear models (GLMs) are provided by \citet{li2018mixtures}. While closed form expressions for the  marginal likelihood are available for Gaussian models \citep{li2018mixtures, bayarri2012criteria}. 

\subsection{The algorithm}\label{sec:alg}

Knowing how to compute the (approximate) marginal likelihoods of models, $\hat p(Y|\M)$, the main challenge in  \eqref{modelposterior} is the computation of the denominator. For most practical situations it is an impossible task to compute the marginal likelihoods for all models within $\mathcal{M}$. Instead, the posterior model probabilities~\eqref{modelposterior} are approximated by
\begin{equation}\label{estmodelposterior}
\hat p(\M|Y) =\frac{ \hat p(Y|\M)\ p(\M)}{ \sum_{\M'\in\mathcal{M}^*}\hat p(Y|\M')\ p(\M')},
\end{equation}
where $\mathcal{M}^*$ is either the set of all models visited by the algorithm or a suitable subset thereof (see Section \ref{subsec:inspect.results} for more details). To this end it is essential that the algorithm visits those models with large values of $p(Y|\M')p(\M')$. Note that the estimate~\eqref{estmodelposterior} allows for embarrassingly parallel computing.

Our model space is complex and typically multimodal due to possibly highly correlated features. Multimodality can be dealt with using a mode jumping MCMC algorithm~\citep[MJMCMC]{hubin2018mode}. However, due to the huge number of potential nonlinear features, only a subset of them (called population) can be considered at a time, leading to the Genetically modified MJMCMC algorithm~\citep[GMJMCMC]{hubin2021flexible}.

The search algorithm is iteratively running through the following two steps:
\begin{enumerate}
\item Search through models from a population of features $\mathcal{P}^t\subset\mathcal{F}$;
\item Based on posterior probabilities of features in $\mathcal{P}^t$, update the population
$\mathcal{P}^t\rightarrow \mathcal{P}^{t+1}$.
\end{enumerate}
Let $I_t \subset \{1,\dots,q\}$ be the set of indices that correspond to the features of the $t$-th population $\mathcal{P}^t$. Then in  step 1, models of the form
\[
Y_i=\beta_0+\sum_{j\in I_t}\gamma_j\beta_jF_j(\bm x) + \epsilon_i
\]
are evaluated through the MJMCMC algorithm, which allows for jumps between modes.
When combined with the “population update” in step 2, this results in the GMJMCMC algorithm, which is used to search through $\mathcal{F}$.
The updates in step 2 are based on the three types of transformations described earlier (\textit{projection}, \textit{modification}, and \textit{interaction}), but also allow for original covariates to be included that are currently not in $\mathcal{P}^t$ (\textit{mutation}).
By default, newly generated features that are linearly dependent on existing features within the population are discarded; this multicollinearity check can be disabled if desired (see Section \ref{SubSec:GAM} for an example).  When generating new nonlinear features for $\mathcal{P}^{t+1}$, a feature $F_j(\bm x)$ from $\mathcal{P}^t$ is selected according to the estimated marginal posterior probability of  $\gamma_j$ from $\mathcal{P}^t$.
The starting population $\mathcal{P}^1$ usually consists of the set of initial covariates $\{F_j(\bm x)=x_j,j=1,...,p\}$, though in some cases  it might also consist of a subset thereof (see Example 3 in Section \ref{SubSec:Linear}).

There are several tuning parameters involved in the algorithm. In Section~\ref{Sec:BasicIntro}, we specifically discuss the size of the population $\mathcal{P}^t$, the number $N$ of MJMCMC iterations performed within step 1, and the number $T$ of outer iterations $t \in \{1,..,T\}$. Additionally, the number of parallel runs can also affect the results.
The detailed algorithmic description is available in \citet{hubin2021flexible}. However, we emphasize that the \pkg{FBMS} package offers a completely new implementation of GMJMCMC that is considerably more coherent and user-friendly than the original version. Further information on all tuning parameters is provided in Appendix~\ref{subset:gen.probs} and Appendix~\ref{subset:gen.params}.

\subsection{Main functions of the FBMS package}

When fitting linear models in R, users can choose between high-level functions inspired by \code{lm}, which allow model specification via formulas, and lower-level functions inspired by \code{lm.fit}, which require explicit input of responses and design matrices. The \pkg{FBMS} package adopts a similar dual-layer design. The high-level \code{fbms} function enables model specification using standard R formulas, making it accessible to a broad range of users. In contrast, lower-level functions such as \code{gmjmcmc} and \code{mjmcmc} offer greater control over the modeling process but require manual specification of responses, covariates, and additional parameters. This layered structure accommodates both novice users and experienced statisticians, balancing flexibility with ease of use.

The general-purpose function \code{fbms} serves as a wrapper for a set of lower-level samplers, currently including \code{mjmcmc}, \code{gmjmcmc}, \code{mjmcmc.parallel}, and \code{gmjmcmc.parallel}. The \code{gmjmcmc} function supports the full BGNLM model class, while \code{mjmcmc} is a restricted version limited to linear predictors. The \code{.parallel} variants enable parallel computation.
In addition, \code{fbms} is designed to support future algorithmic extensions of the \pkg{FBMS} package, such as for example evolutionary variational Bayes \citep{sommerfelt2024evolutionary}. For most users, \code{fbms} will be the preferred entry point due to its simplicity and flexibility. Therefore, we focus here on describing the syntax of \code{fbms}, which internally dispatches to \code{mjmcmc}, \code{gmjmcmc}, or their parallel versions as needed.

%
\section[Package]{The \pkg{FBMS} Package: Getting started}\label{Sec:BasicIntro}
%

This section provides a first introduction to the \pkg{FBMS} package. We will focus on how to fit nonlinear models for a metric outcome variable with the \code{fbms} function. We will start with default settings and then slowly introduce different ways to modify the algorithmic performance of GMJMCMC.

\subsection[First Example]{First Example}\label{SubSec:FirstExample}

Our starting point is the first example of the \code{vignette} which comes with the \pkg{FBMS} package.
It considers astronomic data from $n = 939$ exoplanets,  including both planet and host star attributes. An older version of this data set with fewer exoplanets was used by \citet{hubin2021flexible} to illustrate the ability of BGNLM  to recover Kepler's law from raw data in the spirit of symbolic regression.

 We start by installing the package from CRAN and loading the dataset \code{exoplanet}.
\begin{CodeChunk}
\begin{CodeInput}
R> #install.packages("FBMS")
R> library(FBMS)
R>
R> data(exoplanet)
\end{CodeInput}
\end{CodeChunk}
The dataset contains the following ten variables:\\

\begin{minipage}[t]{0.48\textwidth}
\begin{itemize}
    \item[$Y$] \dots "semimajoraxis"
    \item[$x_1$] \dots "mass"
    \item[$x_2$] \dots "radius" 
    \item[$x_3$] \dots "period" 
    \item[$x_4$] \dots "eccentricity"
\end{itemize}
\end{minipage}
\begin{minipage}[t]{0.48\textwidth}
\begin{itemize}
    \item[$x_5$] \dots "hoststar\_mass"
    \item[$x_6$] \dots "hoststar\_radius"          \item[$x_7$] \dots "hoststar\_metallicity"
    \item[$x_{8}$] \dots "hoststar\_temperature"   
    \item[$x_{9}$] \dots "binaryflag" 
\end{itemize}
\end{minipage}

\ \\

The original data have been taken from the Open Exoplanet Catalogue \cite{exocat}. For a detailed description of the different variables see \cite{rein2012proposal}.  

An approximate version of Kepler's third law can be  formulated as     
\begin{equation}
Y \approx K\left(x_3^2 \times x_5 \right)^{1/3}\; , \label{3kepl1}    
\end{equation}
where the constant $K$ includes the gravitational constant $G$ and some normalizing constant for the mass of the host star. We want to recover this law by fitting a nonlinear model with semimajoraxis
as outcome and the other 9 variables as potential predictors. 
 
The \code{exoplanet} dataset from the \pkg{FBMS} package has the outcome variable in its first column and includes one categorical predictor variable \code{binaryflag}. The \code{fbms} function supports the direct use of factor variables as predictors, when the model is specified using the standard R formula syntax, as in \code{lm} or \code{glm}.

To illustrate how to perform predictions with the \pkg{FBMS} package, we divide the data into a training set consisting of the first 500 planets and a test set containing the remaining observations\footnote{In practice one would use a random split of data but for the purpose of reproducibility we make this deterministic split, specifically because the database is regularly updated.}.  

\begin{CodeChunk}
\begin{CodeInput}
R> train.indx <- 1:500
R> df.train = exoplanet[train.indx, ]
R> df.test = exoplanet[-train.indx, ]
\end{CodeInput}
\end{CodeChunk}

Next, we specify the set of nonlinear functions used to define the feature space $\mathcal{F}$  via modification and projection operators. Table~\ref{Tab:functions} lists all functions currently provided by the \pkg{FBMS} package. In addition, users may include any number of custom made functions, with the only restriction being that they must not return infinite values or \code{NaN} for any finite arguments. There is no requirement for user-defined functions to be differentiable or even continuous, as the package relies on gradient-free optimizers or sampling-based methods for parameter estimation.

\begin{table}[htb!]
\begin{center}
\caption{List of functions available for nonlinear transformations in the \pkg{FBMS} package. The upper-left section of the table includes classical activation functions commonly used in neural networks, along with the logical NOT function, which should be applied only to Boolean predictor variables (see Example 8 in Section~\ref{Sec:Priors}). The upper-right section contains more general functions. The lower half of the table pertains to fractional polynomials (see Section~\ref{SubSec:FracPol}).}
\label{Tab:functions}
\ \\
\begin{tabular}{ll|ll}
\hline
name & function & name & function\\
\hline
sigmoid & $1 / (1 + \exp(-x))$  & sqroot & $|x|^{1/2}$  \\
relu & $\max(x,0)$  &  troot & $|x|^{1/3}$ \\
nrelu & $ \max(-x,0)$  &  sin\_deg & $\sin(x/180*\pi)$  \\
hs & $x > 0$   &  cos\_deg & $\cos(x/180*\pi)$  \\
nhs & $x < 0$ &  exp\_dbl & $\exp(-|x|)$  \\
gelu & $x \Phi(x)$  &  gauss & $\exp(-x^2)$ \\
ngelu & $-x \Phi(-x)$  &  erf & $2  \Phi(\sqrt{2}x) - 1$ \\
not & logical NOT  &  arcsinh & asinh$(x)$ \\
\hline
pm2 & $x^{-2}$    & p0pm2 & $\log(|x|)\ x^{-2}$ \\
pm1 & $\mbox{sign}(x) |x|^{-1}$ &   p0pm05 & $\log(|x|)\ |x|^{-0.5}$  \\
pm05 & $|x|^{-0.5}$   & p0p0 & $\log(|x|)^2$  \\
p0 & $\log(|x|)$    & p0p05 & $\log(|x|)\ |x|^{0.5}$  \\
p05 & $|x|^{0.5}$  &  p0p1 & $\log(|x|)\ x$  \\
p2 & $x^{2}$    &  p0p2 & $\log(|x|)\ x^2$\\
p3 & $x^{3}$    & p0p3 & $\log(|x|)\ x^3$ \\
\hline
\end{tabular}
\end{center}
\end{table}

In our example, we use five functions provided by the \pkg{FBMS} package. For illustrative purposes, we additionally include the custom function $\mbox{to3}(x) = x^3$, even though this transformation is already available in the package under the name \code{p3}. To explicitly recover Kepler's third law, it is essential to include the cube root function $\mbox{troot}(x) = x^{1/3}$.
The names of the functions to be passed to the \code{gmjmcmc} function are collected in a character vector.

\begin{CodeChunk}
\begin{CodeInput}
R> to3 <- function(x) x^3
R> transforms <- c("sigmoid","sin_deg","exp_dbl","p0","troot","to3")
\end{CodeInput}
\end{CodeChunk}


\subsubsection{Single chain analysis with default settings}

We are now ready to use the \code{fbms} function to fit nonlinear models to the \code{exoplanet} data set. To begin, we will use the default parameter settings. Later, we will show how to adjust these settings to improve convergence. While it is generally advisable to use the parallel version \code{gmjmcmc.parallel}  to take advantage of multiple threads, we will start with a single-threaded run for the sake of simplicity.

\begin{CodeChunk}
\begin{CodeInput}
R> # single thread analysis
R> result.default <- fbms(formula = semimajoraxis ~ 1 + . , data = df.train,
R>                        method = "gmjmcmc", transforms = transforms)
          
\end{CodeInput}
\end{CodeChunk}

Based on the value of the \code{method} argument, the \code{fbms} function invokes \code{gmjmcmc} using default settings and standard R formula syntax. In this example, \code{semimajoraxis} is specified as the outcome variable, while all other variables in the \code{df.train} data frame (provided via the \code{data} argument) are treated as input covariates for constructing nonlinear features.

If the \code{formula} argument is omitted—as in the calls below—the default behavior is to treat the first variable in \code{df.train} as the outcome and the remaining variables as predictors. The \code{transforms} argument is used to specify the set of nonlinear functions applied for feature generation. All other arguments of \code{gmjmcmc} are left at their default values.
Unless explicitly stated otherwise, \code{gmjmcmc} assumes a Gaussian response, an identity link function, and Zellner’s g-prior for the regression coefficients. Further details will be provided later.

We now discuss some of the output produced by \code{gmjmcmc} when \code{verbose = TRUE}, which is also the default setting. The algorithm begins with 100 MJMCMC iterations that include only the original input covariates as linear predictors ($t=1$).

\begin{CodeChunk}
\begin{CodeOutput}  
New best crit in cur pop: 698.713835722911 
New best crit in cur pop: 705.633782823956 
New best crit in cur pop: 707.690289643544 
New best crit in cur pop: 707.718019438353 
New best crit in cur pop: 710.793718114534 
 |========================================|
Population 1 done.
Current best crit: 710.793718114534 
Feature importance:
 ###########!#################| mass 
            !               ##| radius 
############!#################| period 
            !               ##| eccentricity 
            !                #| hoststar_mass 
            !                 | hoststar_radius 
            !               ##| hoststar_metallicity 
            !                #| hoststar_temperature 
            !               ##| binaryflag 
\end{CodeOutput}
\end{CodeChunk}

The criterion presented at the beginning of the output is the logarithm of the model posterior probabilities (\ref{modelposterior}) up to a constant. The first criterion with approximate value 698.71 corresponds to the first model visited by the algorithm. Within the 100 MJMCMC iterations of the first round there were 4 improvements of that model. 
Among the linear models from the initial MJMCMC iterations,  \code{period} has the largest estimated marginal posterior probability, followed by \code{mass}. 

In the next step, GMJMCMC updates the population of models by using the four different operators: interaction (with 40\% probability), modification (with 40\% probability), nonlinear projection (with 10\% probability) and mutation (with 10\% probability).

\begin{CodeChunk}
\begin{CodeOutput}
Replaced feature radius with (period*mass) 
Replaced feature eccentricity with sigmoid(mass) 
Replaced feature hoststar_mass with (period*period) 
Replaced feature hoststar_radius with sin_deg(mass) 
Replaced feature hoststar_metallicity with troot(eccentricity) 
Replaced feature hoststar_temperature with sin_deg(sigmoid(mass)) 
Replaced feature binaryflag with (mass*mass) 
Added feature exp_dbl(radius) 
Added feature sigmoid(1+1*hoststar_radius+1*radius+1*period+
                      1*mass+1*binaryflag+1*hoststar_mass+1*exp_dbl(radius)) 
Added feature p0(mass) 
Added feature sin_deg(hoststar_mass) 
\end{CodeOutput}
\end{CodeChunk}

Seven features from the initial population of linear terms are replaced by nonlinear features like for example 
\code{period*mass} or \code{sigmoid(mass)}. Additionally the population has been increased by adding four new features\footnote{See the description of the parameter \code{params\$feat\$pop.max} in Appendix \ref{subset:gen.params}}. This increase of the population size only happens per default after the first MJMCMC round, when switching from linear to nonlinear features. The algorithm then performs a second MJMCMC round using the features from the new population:

\begin{CodeChunk}
\begin{CodeOutput}
New best crit in cur pop: 696.607863376437 
New best crit in cur pop: 709.037079573282 
New best crit in cur pop: 715.764079236183 
New best crit in cur pop: 729.68070968453 |
New best crit in cur pop: 900.04247176188 |
New best crit in cur pop: 900.430134306001 
 |========================================|
Population 2 done.
Current best crit: 900.430134306001 
Feature importance:
            !     ############| mass 
 ###########!#################| (period*mass) 
############!#################| period 
            !                 | sigmoid(mass) 
############!#################| (period*period) 
############!#################| sin_deg(mass) 
            !                 | troot(eccentricity) 
            !                 | sin_deg(sigmoid(mass)) 
 ###########!#################| (mass*mass) 
            !                 | exp_dbl(radius) 
            !                 | sigmoid(1+1*hoststar_radius+1*radius+1*period+1*mass+1*binaryflag+1*hoststar_mass+1*exp_dbl(radius)) 
            !                 | p0(mass) 
            !                 | sin_deg(hoststar_mass) 
\end{CodeOutput}
\end{CodeChunk}

\code{period} still has a fairly large marginal posterior but there are already several strong nonlinear competitors, in particular \code{period*mass}, \code{period}$^2$, $\sin$(\code{mass}) and \code{mass*mass}.  

From now on the algorithm iteratively continues  to update populations with newly generated features (interaction, modification, nonlinear projection) or bringing back initial covariates (mutation) and then runs MJMCMC on each new population, which then yields the final output:

\begin{CodeChunk}
\begin{CodeOutput}
Population 10 done.
Current best crit: 1010.03893076674 
Feature importance:
 ###########!#################| troot(period) 
            !                 | radius 
############!#################| period 
            !                 | p0(period) 
            !                 | p0((period*eccentricity)) 
            !                 | sigmoid((period*eccentricity)) 
 ###########!#################| (troot(period)*sigmoid((period*eccentricity))) 
 ###########!#################| (troot(period)*period) 
            !                 | (sigmoid(troot(hoststar_radius))*sigmoid(troot(hoststar_radius))) 
            !                 | (period*eccentricity) 
            !                 | (radius*binaryflag) 
            !                 | to3((troot(period)*sigmoid((period*eccentricity)))) 
            !                 | (hoststar_radius*troot(period))
\end{CodeOutput}
\end{CodeChunk}

The feature \code{period} still has the highest posterior probability, followed by three other nonlinear features: \code{period}$^{1/3}$, \code{period}$^{4/3}$, and \code{sigmoid(period * eccentricity)} \code{period}$^{1/3}$. However, this result does not correspond to the correct physical relationship. It suggests that the GMJMCMC algorithm, with its default settings, failed to explore the highest-probability regions of the model space, which we would expect to align with the true physical law. In the next section, we demonstrate how to improve model performance by increasing the number of GMJMCMC populations \code{P} and running more MJMCMC iterations for each population (i.e., increasing \code{N} and \code{N.final}).

\subsection{Single chains analysis with more iterations} \label{Subsec:MoreIter}

Before we run the previous example with a larger number of iterations both for the genetic algorithm and for the MJMCMC chains let us have a closer look at all the different algorithmic options that are available. 
The \code{gmjmcmc} function, when invoked via \code{fbms}, accepts fifteen arguments, of which at least the first two must be specified. The purpose of each argument is described below.
\begin{enumerate}
    \item \code{data}: A data frame or matrix containing the data to be used for model fitting. If the outcome variable is in the first column of the data frame, the \code{formula} argument in \code{fbms} can be omitted, provided that all other columns are intended to serve as input covariates.\footnote{if gmjmcmc is called on its own, then data has to be provided as $y$ and $x$ like in \code{lm.fit} and  \code{glm.fit}.} 

    \item  \code{transforms}: A character vector including the names of the nonlinear functions to be used by the modification and the projection operator. 

\item  \code{P}:   The number of population iterations for GMJMCMC. 
The default value is \code{P} = 10, which was used in our initial example for illustrative purposes. However, a larger value, such as \code{P} = 50, is typically more appropriate for most practical applications.

   \item  \code{N}:   The number of MJMCMC iterations per population. The default value is \code{N} = 100, which was used in the first example for illustration; however, for real applications, a larger value such as \code{N} = 1000 or higher is often preferable.

   \item  \code{N.final}: The number of MJMCMC iterations performed for the final population. Per default one has \code{N.final = N}, which was set to 100 in the first example, but for practical applications, a much larger value (e.g., \code{N.final} = 1000) is recommended. Increasing \code{N.final} is particularly important if predictions and inferences are based solely on the last population (an option discussed later in this paper in detail).

 \item  \code{probs}: A list of various probability vectors used by the GMJMCMC algorithm. These are generated by the \code{gen.probs.gmjmcmc} function, as described in Section~\ref{subset:gen.probs}. Of primary importance is the parameter \code{probs.gen}, which is used to define the probabilites of the differemt operators in the feature generation process of the GMJMCMC algorithm. It can be used to select specific classes of nonlinear models, as illustrated in the examples in Section~\ref{Sec:Metric}. Per default \code{probs.gen} gives probabilities of 0.4 to interactions and modifications and probabilites of 0.1 to projections and mutations. The idea behind this choice is to encourage the generation of more interpretable nonlinear features.

   \item  \code{params}: A list of various parameter vectors used by the GMJMCMC algorithm. This list is generated by the \code{gen.params.gmjmcmc} function, as described in Section~\ref{subset:gen.params}.

   \item \code{loglik.pi}: This argument specifies the function which is used to calculate the marginal log-posterior of the model up to a constant and the logarithm of the model prior, that is $\log(\M|Y) = const + \log p(Y|\M) + \log p(\M)$, see also  equation (\ref{modelposterior}). Typically the marginal likelihood is of the form
   \[
p(Y|\M)=\int_{\bm\beta}p(Y|\M,\bm\beta)p(\bm\beta|\M)d\bm\beta,
  \] when the variance of the response is fixed.
In this section we will always fit a Gaussian model with Zellner's g prior (see Section \ref{SubSec:PriorSpec}) which gives the explicit expression (\ref{mloglik.gprior}) for the integral above. This  choice is provided by the default option. In Section \ref{Sec:Priors}  and Section \ref{Sec:Extensions} we will discuss other types of  priors and models, respectively.   

\item  \code{loglik.alpha}: For the nonlinear projection operator the features $F_j(\bm x)$ depend on additional parameters $\bm\alpha$, 
\[
   F_j(\bm x,\bm\alpha)=g(\alpha_{j,0}+\sum_k\alpha_{j,k}F_k(\bm x,\bm\alpha_k)
   \]
   where the sum is taken over a subset of features from the previous population. The argument \code{loglik.alpha} becomes relevant only if the parameter vector $\bm\alpha_j$ is actually estimated.
The simplest method, corresponding to \code{params$feat$alpha = "unit"}, sets all $\alpha$ values to 1. This is the fastest approach and the default setting, although it is certainly often not the most desirable option.
Four alternative methods were discussed by \citet{hubin2021flexible}, but only two
of them are currently implemented in the \pkg{FBMS} package,  strategy 3 (\code{params$feat$alpha = "deep"}; first transform, then optimize across all layers) and strategy 4 (\code{params$feat$alpha = "random"}; the fully Bayesian approach).

 \item \code{mlpost_params}:    
 This is used to provide extra parameters for user-defined priors and marginal likelihoods. Specific examples are given in Section \ref{Sec:Priors}  and Section \ref{Sec:Extensions}.
 
 \item \code{beta_prior}:  Different priors for the regression coefficients can be chosen. All options are listed in Table \ref{tab:priors} of Section \ref{SubSec:PriorSpec} and examples are given in Section \ref{Sec:Priors}.   

\item \code{intercept}: A Boolean indicator specifying whether an intercept should be added to the design matrix. By default, the intercept is included, and no variable selection is performed on it.

 \item \code{fixed}:  Specifies the number of leading columns in the design matrix that will always be included in the model. By default, this value is zero. These covariates will not be used for feature generation. For custom functions they will always be passed to the marginal likelihood estimator.

   \item  \code{sub}: A Boolean indicator that determines whether the full likelihood is used (default) or if the analysis is based on a subsampling \citep{lachmann2022subsampling} or more generally stochastic approach (when the marginal likelihood can be improved at each visit of the same model). An example is provided in Section~\ref{SubSec:sub}.

   \item  \code{verbose}: A Boolean indicator specifying if messages should be printed or not. Per default it is true for \code{gmjmcmc} but false for its parallel version.
\end{enumerate}

We now repeat the single-thread analysis using $P = 50$ populations instead of the default $P = 10$. Additionally, we set \code{N} = 1000 for the MJMCMC iterations throughout the algorithm and \code{N.final} = 5000 for the MJMCMC iterations on the final population. Note that these changes will substantially increase the algorithm’s runtime.

\begin{CodeChunk}
\begin{CodeInput}
R> result.P50 <- fbms(data = df.train, method = "gmjmcmc", 
R>                 transforms = transforms, P = 50, N = 1000, N.final = 5000)
\end{CodeInput}
\end{CodeChunk}

The final lines of the corresponding output from \code{gmjmcmc} read as follows:

\begin{CodeChunk}
\begin{CodeOutput}
Population 50 done.
Current best crit: 1060.66724549699 
Feature importance:
            !                 | to3(hoststar_radius) 
            !                 | (eccentricity*period) 
            !                 | exp_dbl(period) 
############!#################| (troot((period*hoststar_temperature))*
                                 troot((period*hoststar_temperature))) 
            !                 | exp_dbl(exp_dbl(hoststar_radius)) 
            !                 | to3(troot(mass)) 
            !                 | (period*radius) 
            !                 | exp_dbl(to3(hoststar_radius)) 
            !                 | (exp_dbl(period)*hoststar_temperature) 
            !                 | sin_deg(hoststar_mass) 
            !                 | (hoststar_mass*(hoststar_radius*troot(mass))) 
            !                 | sin_deg(exp_dbl(to3(hoststar_radius))) 
            !                 | (sin_deg(exp_dbl(to3(hoststar_radius)))*period) 
\end{CodeOutput}
\end{CodeChunk}

While still not correct, the feature \code{(period*hoststar_temperature)}$^{2/3}$ is close to the feature \code{period}$^{2/3}$\code{*hoststar_temperature}$^{1/3}$, which is highly correlated with the correct solution representing Kepler's law. This illustrates how increasing the number of iterations can improve convergence. In the next step, we will demonstrate how running multiple parallel chains can further enhance model identification.

\subsection{Analysis with multiple chains}\label{subset:multiple}

MCMC algorithms (even MJMCMC) may struggle with multimodal posterior distributions, especially in nonlinear settings. A common strategy to improve exploration of the model space is to run multiple chains with different initial values. In the \pkg{FBMS} framework, this idea of embarrassingly parallel execution is implemented via the function \code{gmjmcmc.parallel}, which allows multiple chains to run in parallel (see also the discussion following equation~\eqref{estmodelposterior}). Using the \code{fbms} function this is invoked via  
\code{method = "gmjmcmc.parallel"}.

The two new key arguments are \code{runs}, which specifies the number of independent Markov chains to execute, and \code{cores}, which indicates the number of CPU cores to be used. Ideally, these should be set equal to fully utilize all available cores and maximize computational efficiency. If \code{runs} exceeds \code{cores}, chains will be scheduled sequentially, increasing runtime. Conversely, setting \code{runs} lower than \code{cores} underutilizes available resources.
Running multiple chains has been shown to improve inference by enabling exploration of a broader set of models and nonlinear interactions, as demonstrated by \citet{hubin2021flexible}. Note that parallel performance may depend on the operating system: under Windows, \code{gmjmcmc.parallel} can be less efficient than under Unix-based systems due to differences in R’s parallelization back end.

We now present results from running the previous example on a Linux server using 40 parallel chains, with \code{runs = 40} and \code{cores = 40} specified. Each chain used the default GMJMCMC settings with $N = 100$ and was run with $P = 25$ populations, that is half the number used in the earlier single-threaded analysis.

\begin{CodeChunk}
\begin{CodeInput}
R> result.parallel <- fbms(data = df.train, method = "gmjmcmc.parallel", 
R>                    transforms = transforms, runs = 40, cores = 40, P = 25)
\end{CodeInput}
\end{CodeChunk}

Given these settings, the parallel analysis was substantially faster than the previous single-threaded run. Despite using fewer iterations per chain it yielded substantially better results, as we will see in the next section. When multiple cores are used, some operating systems and IDEs (like RStudio) may suppress real-time console output from the parallel version of \code{gmjmcmc}, while other IDEs (like PyCharm) may mess output from different chains up in a single console. To avoid unnecessary overhead from invisible or messy output, we recommend setting \code{verbose = FALSE}, which is also the default for \code{gmjmcmc.parallel}. We will next introduce several useful functions for summarizing and interpreting the results obtained with \code{fbms}.


\subsection{Inspection of Results} \label{subsec:inspect.results}

The \pkg{FBMS} package provides methods for \code{summary}, \code{plot}, and \code{predict}, which work seamlessly with both the single-threaded and parallel versions of the algorithm.
To illustrate their use, we compare the outputs from the two single-threaded analyses (\code{result.default} and \code{result.P50}) with the parallel run \code{result.parallel}.
Additionally, when multiple chains have been executed, diagnostic plots can be generated using the \code{diagn_plot} function to assess convergence and model exploration across chains.
 
\subsubsection{Summary}

\begin{CodeChunk}
\begin{CodeInput}
R> summary(result.default)
\end{CodeInput}
\end{CodeChunk}

gives the following output:

\begin{CodeChunk}
\begin{CodeOutput}
Best   population: 8  log marginal posterior: 1024.187 

           feats.strings marg.probs
1                 period   1.000000
2 (troot(period)*radius)   1.000000
3 (troot(period)*period)   1.000000
4  (period*eccentricity)   1.000000
5          troot(period)   0.998574
\end{CodeOutput}
\end{CodeChunk}

By default, the reported posterior probabilities of features are based on the best population, that is the one which includes the model with the highest marginal posterior. Alternative options include \code{pop = "last"} to use the final population, and \code{pop = "all"} to compute posterior probabilities across all populations. The last option can become quite time consuming depending on the number of visited models.

Complex nonlinear features can sometimes be difficult to interpret when expressed using the original variable names. To improve readability, the \code{labels} argument of the \code{summary} function can be used to provide more user-friendly output, as demonstrated below.

\begin{CodeChunk}
\begin{CodeInput}
R> summary(result.default, pop = "all", 
R>         labels = paste0("x",1:length(df.train[,-1])))
\end{CodeInput}
\end{CodeChunk}

\begin{CodeChunk}
\begin{CodeOutput}
Best   population: 8  thread: 1  log marginal posterior: 1024.187 

   feats.strings marg.probs
1             x3  1.0000000
2 (troot(x3)*x3)  0.9999999
3        (x3*x4)  0.9999993
4      troot(x3)  0.9988126
5 (troot(x3)*x2)  0.8326640
\end{CodeOutput}
\end{CodeChunk}

Here \code{x3} corresponds to \code{period} and 
\code{x4} corresponds to \code{radius}. The posterior probabilities based on all populations are slightly different from those based on the best population.

Features are presented  in the order of their posterior probabilities, where per default features  with posterior larger than $10^{-4}$ are listed. The \code{summary} function has the argument \code{tol} with which this cutoff can be adapted. We illustrate this by presenting posterior probabilities for \code{result.P50} using the last population.

\begin{CodeChunk}
\begin{CodeInput}
R> summary(result.P50, pop = "last", 
R>         labels = paste0("x",1:length(df.train[,-1])))
\end{CodeInput}
\end{CodeChunk}

\begin{CodeChunk}
\begin{CodeOutput}
Best   population: 24  log marginal posterior: 1060.667 
Report population: 50  log marginal posterior: 1060.667 

                    feats.strings   marg.probs
1 (troot((x3*x8))*troot((x3*x8))) 1.0000000000
2                         to3(x6) 0.0008679760
3                         (x4*x3) 0.0004111358
4                         (x3*x2) 0.0002485837
5                     exp_dbl(x3) 0.0002290707
6                     sin_deg(x5) 0.0002264469
\end{CodeOutput}
\end{CodeChunk}

When setting \code{tol = 0.01} only one relevant nonlinear feature with high posterior probability remains:

\begin{CodeChunk}
\begin{CodeInput}
R> summary(result.P50, pop = "last", tol = 0.01,
R>         labels = paste0("x",1:length(df.train[,-1])))
\end{CodeInput}
\end{CodeChunk}

\begin{CodeChunk}
\begin{CodeOutput}
Best   population: 24  log marginal posterior: 1060.667 
Report population: 50  log marginal posterior: 1060.667 

                    feats.strings marg.probs
1 (troot((x3*x8))*troot((x3*x8)))          1
\end{CodeOutput}
\end{CodeChunk}

Finally, we look at the results from the parallel version. 

\begin{CodeChunk}
\begin{CodeInput}
R> summary(result.parallel)
\end{CodeInput}
\end{CodeChunk}

where we obtain

\begin{CodeChunk}
\begin{CodeOutput}
Best   population: 10  thread: 19  log marginal posterior: 1078.438 

                           feats.strings   marg.probs
1 troot(((hoststar_mass*period)*period)) 0.9999561337
2               sigmoid(hoststar_radius) 0.0002209012
\end{CodeOutput}
\end{CodeChunk}

The parallelized algorithm  \code{gmjmcmc.parallel} run with 40 threads actually succeeded to recover Kepler's third law. The corresponding nonlinear term has a posterior of more than 99\% and is the only relevant nonlinear feature. 
Computing posteriors based on all populations yields exactly the same result, but it takes already 16 seconds to get the summary.

\subsubsection{Plots}
\begin{figure}[ht!] 
    \centering
\begin{minipage}[c]{0.43\textwidth}

\includegraphics[width=0.9\textwidth]{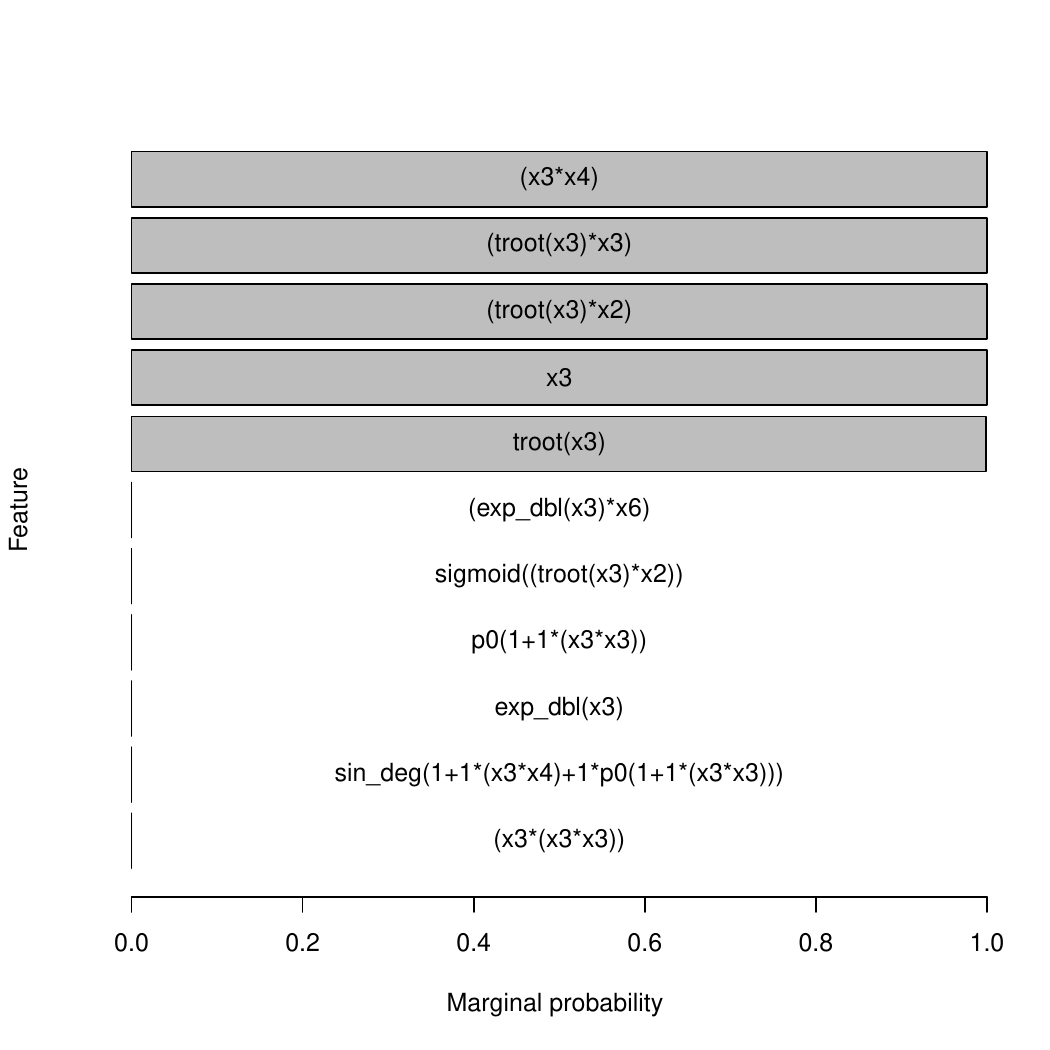}
\end{minipage}
\begin{minipage}[c]{0.43\textwidth}

\includegraphics[width=0.9\textwidth]{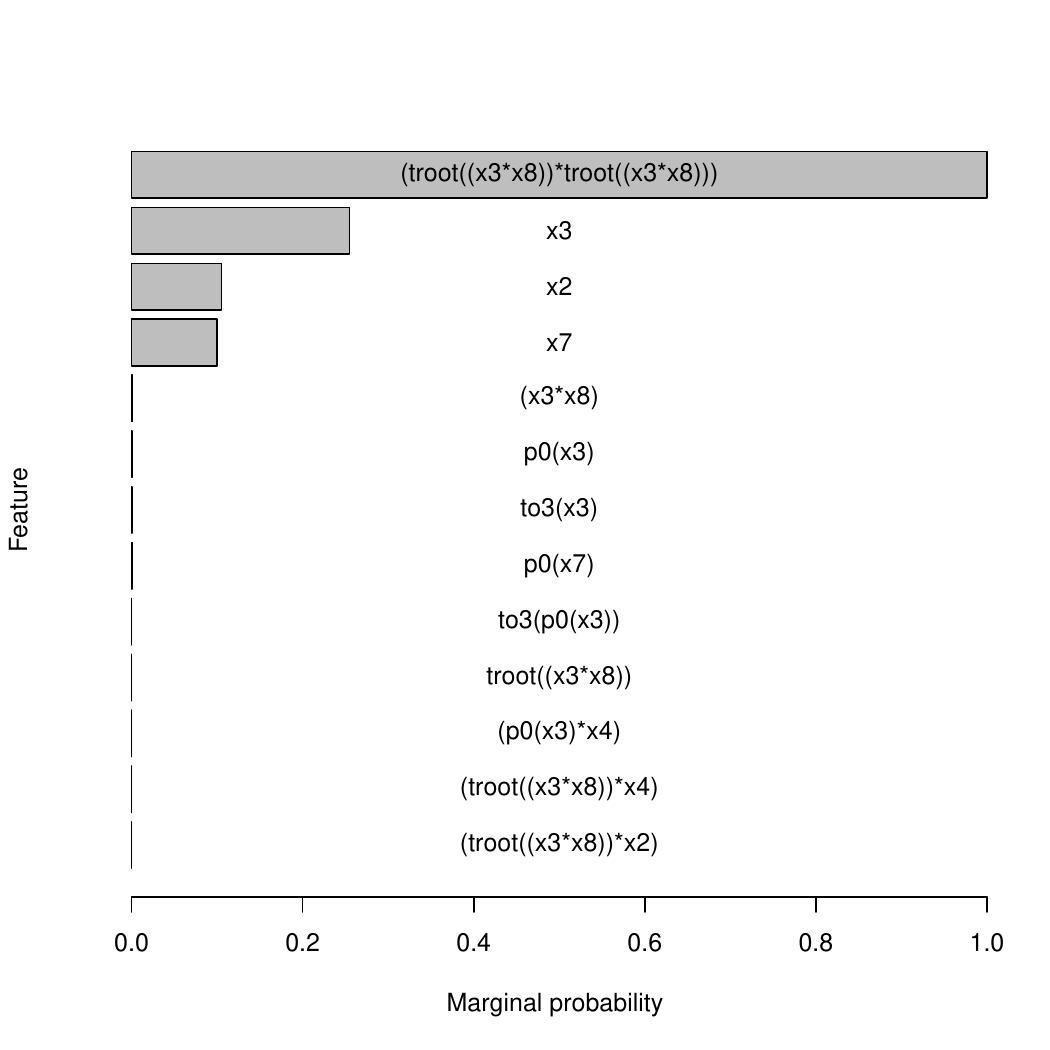}
\end{minipage}    
  
\caption{Results from single thread analysis illustrated with the \code{plot} function. }\label{fig:1}
\end{figure}

Apart from \code{summary} there is also the \code{plot}  command to get information about the results from \code{fbms}. The following code gives the two graphs from Figure \ref{fig:1}, where by default the best population is used to compute posterior probabilities of features. 

\begin{CodeChunk}
\begin{CodeInput}
R> plot(result.default)
R> plot(result.P50)
\end{CodeInput}
\end{CodeChunk}

Note that for \code{result.P50} the plotted results differ slightly from  those obtained with \code{summary}, where we had used not the best but the last population to perform model averaging. The plot for \code{result.parallel} may not be visually appealing if too many features are included. One can limit the number of features displayed, for example, to 6:\begin{CodeChunk}
\begin{CodeInput}
R> plot(result.parallel, count = 6)
\end{CodeInput}
\end{CodeChunk}

The resulting plot (in this example) highlights the correct feature with a posterior probability close to 1, along with five others that have very low posteriors.
The \code{plot} function provides essentially the same information as the \code{summary} output, but in a more visually structured form. By default, it shows features from the best population (\code{pop = "best"}) ordered by their posterior probabilities. As with \code{summary}, the population can alternatively be set to \code{"last"} or \code{"all"} using the \code{pop} argument.

\subsubsection{Predict}

Another useful function for working with results from \code{gmjmcmc} is \code{predict}, which takes new data as input and uses Bayesian model averaging to generate predictions based on the posterior modes of individual models.  Users can optionally specify a link function, which will become important in case of exponential family response models (see Section \ref{Sec:Extensions}).

Just like \code{summary} and \code{plot}, the \code{predict} function supports three different modes by setting the \code{pop} argument: using models from the best population (\code{pop = "best"} — the default), from all populations (\code{pop = "all"}), or only from the last population (\code{pop = "last"}). Prediction using models from all populations can be computationally intensive and require substantial memory. Therefore, it may be preferable to use predictions based on the best or last population in practice. 
Predictions from individual models are computed using posterior modes of the parameters and then aggregated using model posterior probabilities.

The object returned by the \code{predict} function contains prediction results for each new data point from each parallel chain. Assuming \code{preds} is the output of a \code{predict} call, aggregated predictions across chains can be found in the list \code{preds$aggr}\footnote{Note that for a single MJMCMC chain, that we will use in the  next example, \code {pred} has these elements directly.} which includes the following components:

\begin{enumerate}
\item \textbf{Mean}: The weighted mean prediction for the new data, calculated based on the models' posterior probabilities.
\item \textbf{Quantiles}: Weighted quantiles of the predictions that provide credible intervals for the predicted values, accounting for model probabilities.
\end{enumerate}

These components enable users to assess not only the overall prediction but also the uncertainty and distribution of predictions across different models and populations (the latter when using \code{pop = "all"}), providing a comprehensive view of the predictive performance of the variable selection procedure. An example will be given in Section \ref{SubSec:GAM}.

We will now generate predictions on the test data and then plot the predicted values against the actual outcome values.

\begin{CodeChunk}
\begin{CodeInput}
R> preds <-  predict(result.default, df.test[,-1])  
R> plot(preds$aggr$mean, df.test$semimajoraxis)

R> preds.P50 = predict(result.P50, df.test[,-1]) 
R> plot(preds.P50$aggr$mean, df.test$semimajoraxis)

R> preds.multi <- predict(result.parallel , df.test[,-1])  
R> plot(preds.multi$aggr$mean, df.test$semimajoraxis)
\end{CodeInput}
\end{CodeChunk}

\begin{figure}[th!]
    \centering
\begin{minipage}[c]{0.48\textwidth}

\includegraphics[width=0.9\textwidth]{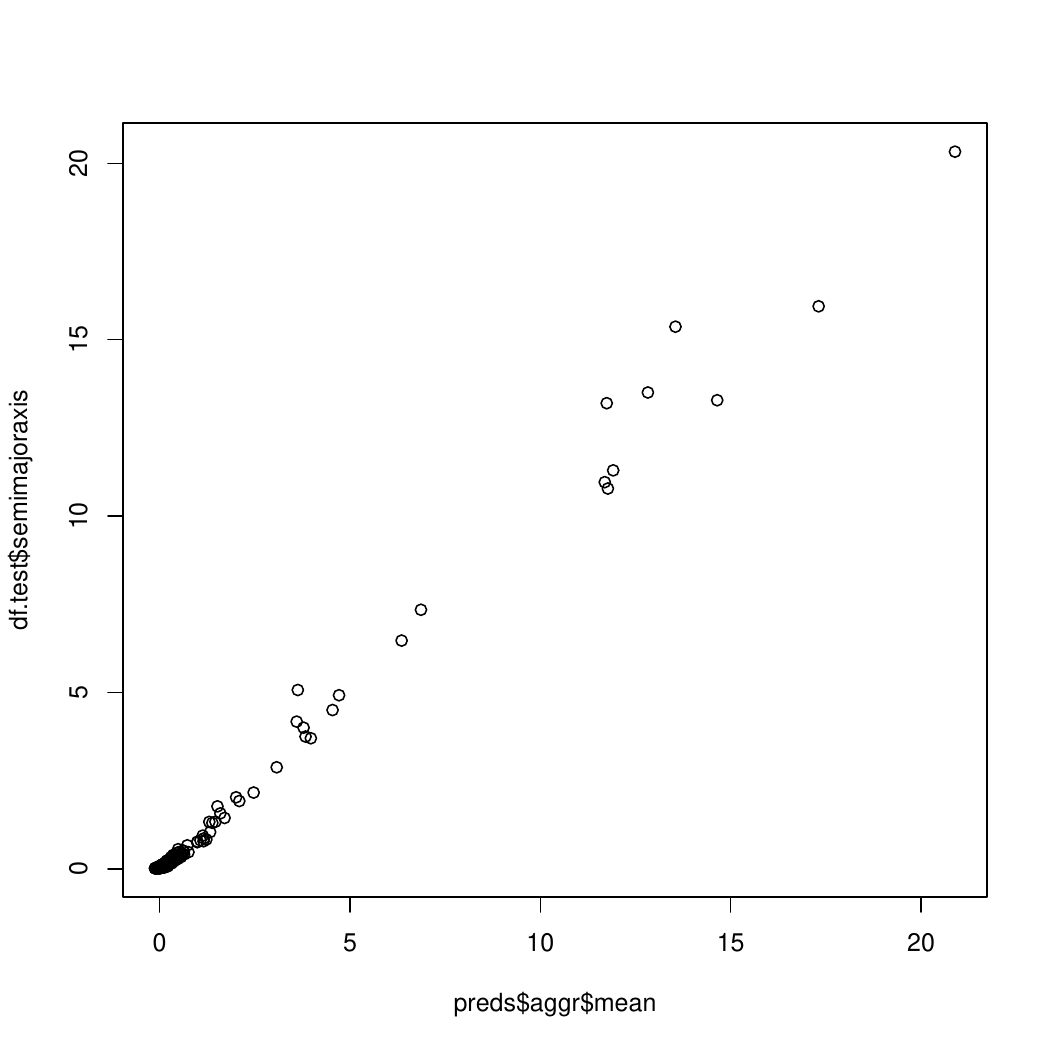}
\end{minipage}
\begin{minipage}[c]{0.48\textwidth}

\includegraphics[width=0.9\textwidth]{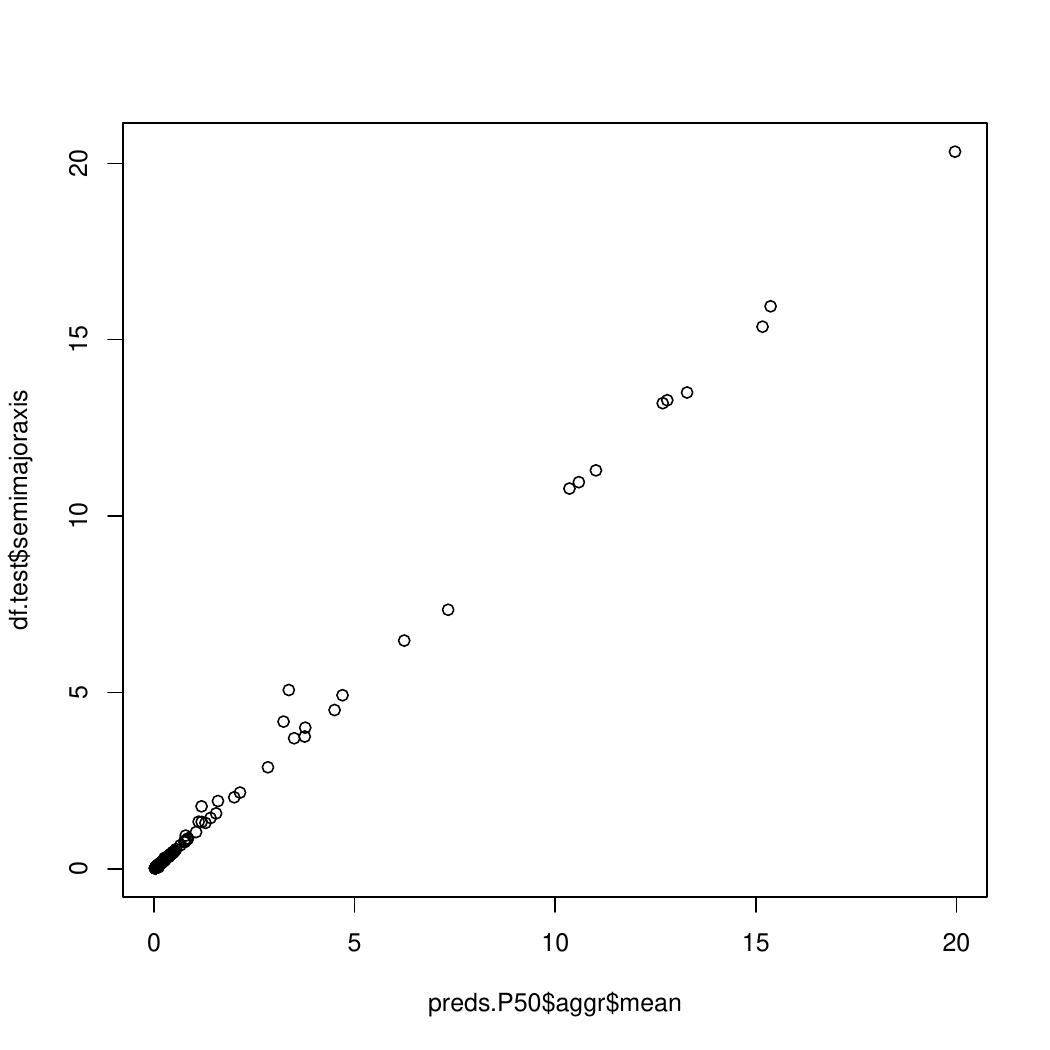}
\end{minipage}
 \caption{Quality of prediction from the two single thread analysis results, the left plot based on the default settings and the right one using larger numbers of iterations as specified for \code{result.P50}.}  \label{fig:2}
\end{figure}

Figure \ref{fig:2} presents the prediction plots from the two single-thread analyses. Visual inspection reveals noticeable differences, which are reflected in the root mean squared errors (RMSE) of 0.20 and 0.12, respectively. Further, out-of-sample prediction based on \code{result.parallel} performs nearly perfectly, achieving an RMSE of 0.02, indicating successful recovery of the data-generating process.

Apart from model averaging, \code{predict} can also make predictions based on the best model (obtained with \code{get.best.model}) or the median probability model selecting features with marginal posterior above 0.5 (obtained with \code{get.mpm.model}), as we will explore in Example 12, Section \ref{SubSec:Survival}.

\subsection{Diagnostic plots for parallel runs}

It is important to note that different chains run with \code{gmjmcmc.parallel} typically produce varying posterior probabilities for each nonlinear feature. The \code{summary} function reports combined posterior probabilities by computing a weighted average of the individual chain results (see \citet{hubin2021flexible} for details). Examining the variation in posterior probabilities across separate chains can provide valuable insight into the stability of the results. Greater variability between chains suggests potential convergence difficulties for GMJMCMC. However, perfect agreement among chains does not necessarily guarantee convergence. It could also indicate that the chains have become trapped in the same or in a similar mode.

The \code{diagn_plot} function provides a useful tool to visualize the stability of solutions obtained from different parallel GMJMCMC chains. It assesses convergence by plotting summary statistics of log posterior values over iterations, similar to DAG score plots used in MCMC-based structure learning for Bayesian networks \citep{JSSv105i09}. Specifically, \code{diagn_plot} shows how a chosen summary statistic (such as the median, mean, minimum, maximum, or variance) of the log posterior evolves as the GMJMCMC populations progress.

For multiple chains, the function calculates a summary value (e.g. the median) of the log posterior within each population by combining results across all chains. It then measures how much this summary varies over time by looking at its fluctuations within a sliding window of recent populations. Using this variation, the function creates confidence intervals that show how stable or variable the summary is as the algorithm progresses. When there is only a single chain, the variation is estimated by looking at how the summary changes within the sliding window over consecutive populations from that chain. The resulting plot shows the summary together with upper and lower confidence bounds, helping to visualize convergence and stability over time. Plot appearance can be customized through standard R graphical options.

Such visualization helps evaluate whether the chains are mixing around a stable solution, indicating that the GMJMCMC algorithm has sufficiently explored the parameter space and that the resulting estimates are reliable. However, even very stable convergence on the plot does not guarantee that a global maximum of the log posterior has been reached. At times, there is the illusion of perfect convergence when the algorithm settles into a strong posterior mode from which it cannot escape. In practice, mixing around local sub-modes without large jumps up or down is often preferable, as it suggests good mixing across populations and thorough exploration of the model space.

Below, we present diagnostic plots for the two previously fitted models, \code{result.default} and \code{result.parallel}. We call \code{diagn_plot} using as summary statistics the maximum of the the log marginal posteriors across populations.  

\begin{CodeChunk}
\begin{CodeInput}
R> diagn_plot(result.default, ylim = c(600,1500), FUN = max)

R> diagn_plot(result.parallel, ylim = c(600,1500), FUN = max)

\end{CodeInput}
\end{CodeChunk}

\begin{figure}[!h]
    \centering
    \includegraphics[width=0.45\linewidth]{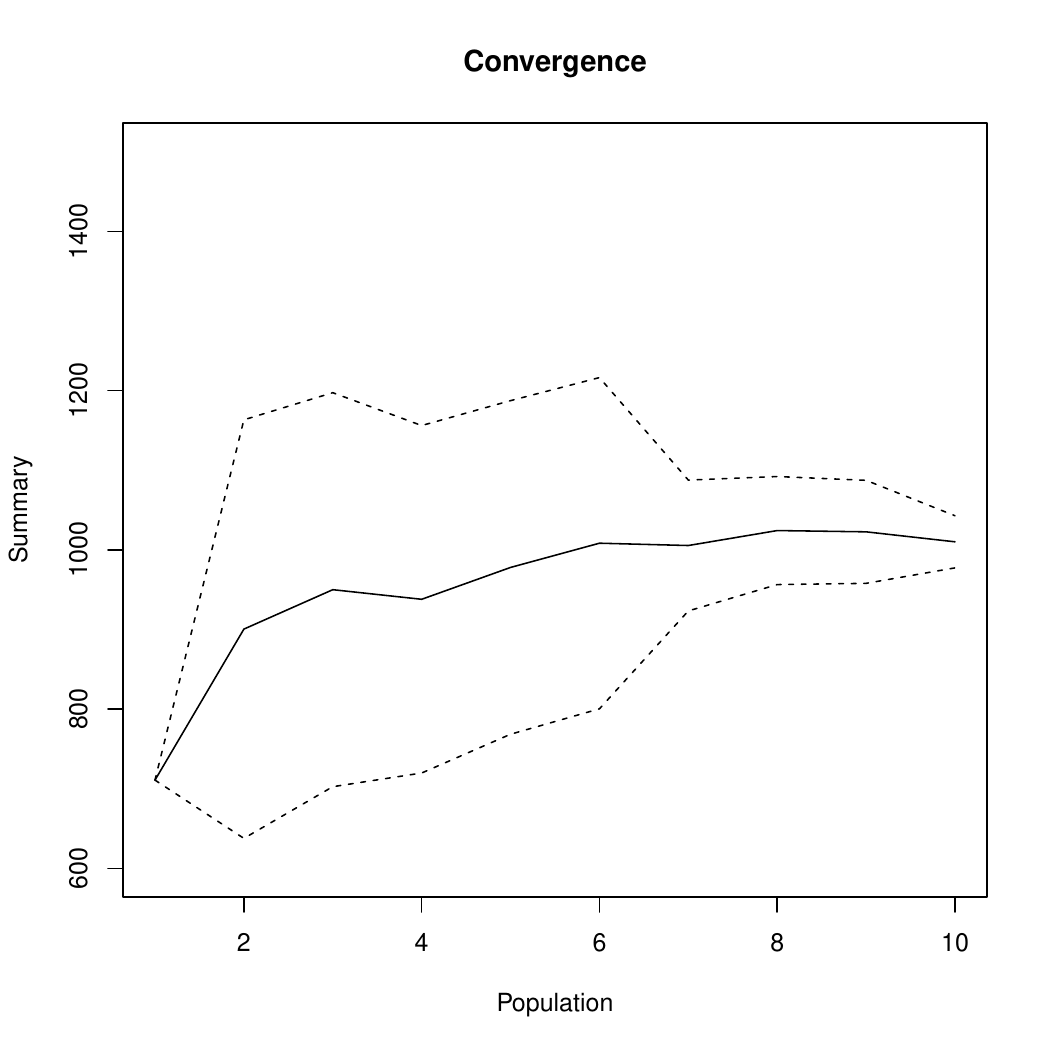}
    \includegraphics[width=0.45\linewidth]{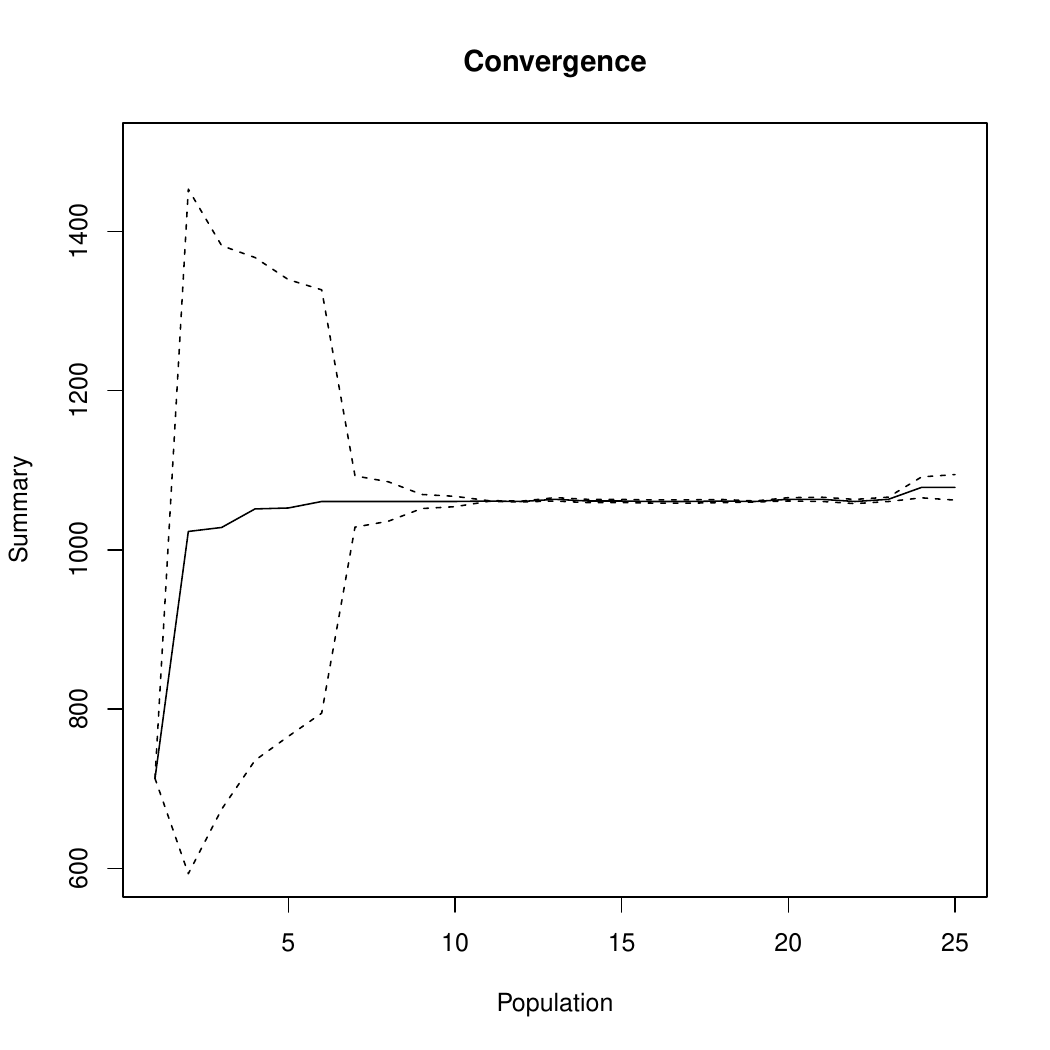}
    \caption{Diagnostic plots for single GMJMCMC chain with default setting (left panel), and parallel version with 40 chains (right panel).}
    \label{fig:diagnplots}
\end{figure}

The resulting plots are shown in Figure \ref{fig:diagnplots}. One can clearly see that the single chain of \code{result.default} has not yet converged after 10 populations. Similarly the parallel chains tend to reach stability only after 10 populations. Note that at the end the parallel chain got a slight improvement, which corresponds to hitting a new mode in one or several of the chains. 

 Improving the convergence of GMJMCMC involves fine-tuning key parameters, such as the mutation and crossover probabilities, and achieving an effective balance between exploration and exploitation to prevent premature convergence to local modes. Maintaining a sufficiently large and diverse population increases the chance of discovering global modes, and running more iterations per population can further aid convergence. Running multiple independent chains in parallel typically enhances robustness, as pooling results across chains yields more reliable posterior estimates. Post-processing tools such as \code{diagn_plot} are valuable for assessing convergence stability, which helps to identify poor mixing, stagnation, or other convergence issues, and provide actionable feedback for adjusting algorithm settings. Restart strategies, such as reinitializing chains or incorporating insights from previous runs, can also help to avoid convergence to suboptimal local optima of the posterior terrain.

When analyzing the \code{diagn_plot}, stable trends in the median (or another summary statistic) and non-diverging confidence intervals indicate improved convergence to either a good local or a global mode. While stable \code{diagn_plot} patterns are encouraging, they do not guarantee that the global mode has been identified. Therefore, sensitivity checks and additional chains are important for assessing the robustness of the inference, much like when using DAG score plots in Bayesian network structure learning. 

This concludes the basic introduction to the \pkg{FBMS} package. The behavior of the \code{gmjmcmc} function is further controlled by two main components: a list of probabilities (\code{probs}) and a list of algorithmic parameters (\code{params}), which are described in detail in Section~\ref{subset:gen.probs} and Section~\ref{subset:gen.params}, respectively.
Default settings for these components can be generated using the functions \code{gen.probs.gmjmcmc} and \code{gen.params.gmjmcmc}, and specific values can then be modified as needed.
In the following section, particular attention will be paid to the choice of  \code{probs$gen}, which determines the probabilities with which the four feature generation operators (\textit{interaction}, \textit{modification}, \textit{nonlinear projection}, and \textit{mutation}) are applied.

%
\section[Metric]{Specific Models for a Metric Response}\label{Sec:Metric}
%

In this section, we explore the wide range of model classes that the \pkg{FBMS} package can accommodate through its flexible feature generation mechanism. We continue to assume a metric outcome with Gaussian error terms and adopt the g-prior for the regression coefficients. However, we now examine different families of features $F_{j}(\bm{x}_i)$ that can be constructed. Variants of linear regression models using alternative priors will be discussed in Section~\ref{Sec:Priors}, while models with different likelihood functions are introduced in Section~\ref{Sec:Extensions}.

\subsection[Linear Model]{Example 2: Linear Bayesian Model Selection}\label{SubSec:Linear}

We begin with the simplest scenario, where models are constructed without any nonlinearities. The \pkg{FBMS} package provides two straightforward ways to achieve this. One option is to set the first three components of \code{probs$gen} to zero. This effectively disables the generation of interactions, modifications, and nonlinear projections within the GMJMCMC algorithm, so that only original (linear) features are used when forming new populations. This setting is particularly useful for high-dimensional variable selection, and we will illustrate its purpose with a specific example in Section \ref{SubSec:GPrior}.

Alternatively, one can directly apply the MJMCMC algorithm as described in \citet{hubin2018mode} by using the \code{mjmcmc} function. This corresponds to the first step of the GMJMCMC algorithm, but the MJMCMC search is now carried out over the entire set of available covariates. In this case, the algorithm is not embedded within a genetic algorithm, and no nonlinear features are generated. As a result, fewer parameters need to be specified to tune the algorithm. These can be conveniently generated using the functions \code{gen.probs.mjmcmc} and \code{gen.param.mjmcmc}.  While the associated probabilities and parameters are discussed as computational details in Appendix \ref{subset:gen.probs} and \ref{subset:gen.params}, standard users of the package typically do not need to modify them.

In this simple example the data generating model is Gaussian, as in equation (\ref{themodeleq}), but does not include any nonlinear features:
\begin{equation}\label{linear.model}
  Y =  \sum_{j=1}^{5} \frac{j}{5} \times x_{j} + \epsilon \; .
\end{equation}
Covariates are simulated as i.i.d. standard normally distributed vectors of length $n = 100$. Out of a total of $p = 20$ covariates, the first $k^* = 5$ are included in the data-generating model, with effect sizes $\beta_j$  ranging between 0.2 and 1. The data is then normalized prior to the analysis:

\begin{CodeChunk}
\begin{CodeInput}
R> y<-scale(y)
R> X<-scale(X)/sqrt(n)
R> df <- as.data.frame(cbind(y, X))
\end{CodeInput}
\end{CodeChunk}

Given the effect sizes, we expect that $x_1$ will be the most difficult to detect, whereas $x_5$ should be relatively easy to identify. We begin by analyzing the simulated data using GMJMCMC with the same set of transformations as in the previous example. All parameters are set to their default values.

\begin{CodeChunk}
\begin{CodeInput}
R> to3 <- function(x) x^3
R> transforms <- c("sigmoid","sin_deg","exp_dbl","p0","troot","to3")
R> 
R> result <- fbms(data = df, method = "gmjmcmc", transforms = transforms)
R> summary(result)
\end{CodeInput}
\end{CodeChunk}

\begin{CodeChunk}
\begin{CodeOutput}
  feats.strings marg.probs
1            X4  1.0000000
2            X3  1.0000000
3            X5  1.0000000
4           X18  0.1845023
\end{CodeOutput}
\end{CodeChunk}

The three linear terms $x_3, x_4, x_5$, which have the largest effect sizes in the data-generating model, are successfully identified. Additionally, $x_{18}$ appears as false positive with a small but positive posterior probability (although it is negligible if one uses the median probability model). No nonlinear terms are included in the model. However, the covariates $x_1$ and $x_2$ are entirely missed. To enhance the performance of GMJMCMC and potentially recover these weaker signals, we increase both the number of iterations (\code{N}) and the number of populations (\code{P}).

\begin{CodeChunk}
\begin{CodeInput}
R> result2 <-  fbms(data = df, method = "gmjmcmc", transforms = transforms, 
                         N = 1000, P = 40)
R> summary(result2, tol = 0.1)
\end{CodeInput}
\end{CodeChunk}

\begin{CodeChunk}
\begin{CodeOutput}
  feats.strings marg.probs
1            X4  1.0000000
2            X3  1.0000000
3            X5  1.0000000
4            X2  0.9536766
5            X1  0.1736528
\end{CodeOutput}
\end{CodeChunk}

With the increased number of iterations and populations, $x_2$ now has a posterior probability close to one, and even $x_1$ appears with a small but positive posterior probability. These results suggest that \code{gmjmcmc} is not prone to overfitting by unnecessarily including nonlinear features thanks to the model prior which explicitly penalizes complexity of the latter. However,  when the data-generating model consists solely of linear terms one might also use  \code{mjmcmc} instead of \code{gmjmcmc}. Here is the simplest call of \code{fbms} providing only the data as argument.

\begin{CodeChunk}
\begin{CodeInput}
R> result.lin <- fbms(data = df)
\end{CodeInput}
\end{CodeChunk}

This will invoke \code{mjmcmc} which is the default method in \code{fbms} and apparently does not require the \code{transforms} parameter. It also has a larger default number of MJMCMC iterations with \code{N} = 1000.

For additive models, summaries of posterior modes are available in the \code{summary} by specifying \code{effects = c(0.5,0.025,0.975)} for the chosen quantiles of the posterior in the space of models, giving output similar to another popular package for Bayesian model averaging \code{BAS} \citep{Clyde:Ghosh:Littman:2010} and to the BMA procedure in popular statistical  software \code{JASP} \url{https://jasp-stats.org}. Note that for some parameters the lower/upper quantiles may become exactly 0 due to the possibility of variable selection.

\begin{CodeChunk}
\begin{CodeInput}
R> summary(result.lin,effects = c(0.5,0.025,0.975))
\end{CodeInput}
\end{CodeChunk}

\begin{CodeChunk}
\begin{CodeOutput}
Best log marginal posterior:  56.2259 

$PIP
   feats.strings marg.probs
1             X4 1.00000000
2             X3 1.00000000
3             X5 1.00000000
4             X2 0.94943379
5             X1 0.16503140
6            X15 0.13499927
...
20            X6 0.03525845

$EFF
   Covariate quant_0.5 quant_0.025 quant_0.975
1  intercept         0           0           0
2         X1         0           0      0.6596
3         X2    1.7034           0      1.7383
4         X3    4.8118      4.7573      4.8867
5         X4    4.9803      4.9109      5.0888
6         X5    4.7744      4.3659      4.8573
   ...
\end{CodeOutput}
\end{CodeChunk}

In this example, \code{mjmcmc} produces results very similar to those obtained with \code{gmjmcmc}, and in some cases, it may even have greater power to detect small linear effects. However, the most significant advantage of \code{mjmcmc} lies in its runtime efficiency. It is considerably faster than \code{gmjmcmc}, even when increasing the number of MJMCMC iterations \code{N}.


\subsection[Interactions]{Example 3: Interactions}\label{SubSec:Interaction}

We now simulate data in a similar way as in Example 2, using a Gaussian model with $n = 100$ observations and $p = 20$ input covariates that are i.i.d. standard normally distributed. However, in this case, the data-generating model includes interaction terms:
\begin{equation}\label{interaction.model}
  Y = 1.2 * x_1 + 1.5 * x_2 * x_3 - x_4 + x_5 - 1.3 * x_4*x_5 + \epsilon \; .
\end{equation}
Thus $x_1$ appears as a pure main effect, $x_2* x_3$ represents a pure interaction effect, and $x_4$ and $x_5$ contribute both as main effect and interaction effect to the data generating model. 
  
  The \code{probs} argument can be used to control the probabilities associated with the different feature-generating operators. We begin by using the \code{gen.probs.gmjmcmc} function to generate a list containing the default values for \code{probs}. To ensure that \code{fbms} includes only interaction terms as nonlinear features, we then set \code{probs$gen = c(1, 0, 0, 1)}.

\begin{CodeChunk}
\begin{CodeInput}
R> transforms <- c("")
R> probs <- gen.probs.gmjmcmc(transforms)
R> probs$gen <- c(1,0,0,1)  #Include interactions and mutations
\end{CodeInput}
\end{CodeChunk}

Running \code{gmjmcmc} with the default values yields again poor posterior estimates, so we do not show the results here. Increasing the number of iterations leads to improved estimation accuracy.

\begin{CodeChunk}
\begin{CodeInput}
R> result2 <- fbms(data = df, method = "gmjmcmc", transforms = transforms, 
R>                 N = 1000, probs = probs, P=40)
R> summary(result2, tol = 0.01)
\end{CodeInput}
\end{CodeChunk}

\begin{CodeChunk}
\begin{CodeOutput}
   feats.strings marg.probs
1        (X4*X5) 1.00000000
2        (X3*X2) 1.00000000
3             X1 1.00000000
4             X4 1.00000000
5             X5 1.00000000
6             X8 0.64579977
7             X9 0.63279027
8             X3 0.27304553
9            X17 0.11213759
10           X18 0.03607342
\end{CodeOutput}
\end{CodeChunk}
In this run, all the main effects and interaction terms are correctly identified. However, $x_8$ and $x_9$ also exhibit relatively high posterior probabilities and might be regarded as false positive detections. In this example, running \code{gmjmcmc} with 40 parallel chains did not lead to a substantial improvement in the results. However, when increasing the sample size $n$ to $1000$ observations, perfect results are already obtained for \code{result2} above.

\subsection[Fractional Polynomials]{Example 4: Fractional Polynomials}\label{SubSec:FracPol}

The class of fractional polynomials was introduced by \citet{royston1994regression} to offer a flexible and standardized approach for modeling nonlinear relationships between continuous covariates and an outcome variable. In particular, first-order fractional polynomials typically include the following seven nonlinear functions: ${x^{-2}, x^{-1}, x^{-1/2}, \log(x), x^{1/2}, x^2, x^3}$. These correspond to the Box-Tidwell transformations \citep{box1962transformation} with powers ${-2, -1, -0.5, 0, 0.5, 2, 3}$, where the power 0 is interpreted as $x^0 = \log(x)$. While fractional polynomials of higher order can be defined, in practice only first- and second-order polynomials are commonly used \citep{royston2008multivariable}.

The lower part of Table~\ref{Tab:functions} displays the fractional polynomials of degree 1 (left column) and degree 2 (right column) as implemented in the \pkg{FBMS} package. Fractional polynomial regression can be viewed as a special case of the BGNLM framework when these functions are included in the set $\mathcal{G}$ and only nonlinear modifications of depth 1 are considered.

\citet{hubin2023fractional} assessed the performance of GMJMCMC in the context of fractional polynomials. One of their examples is based on data from the ART study provided by \citet{royston2008multivariable}. This dataset originates from a breast cancer study and includes 6 continuous and 4 categorical predictor variables. Two of the categorical variables ($x_4$ and $x_9$) have three categories each and are encoded using two dummy variables, respectively. Following \citet{hubin2023fractional}, we simulate a response variable $Y$ using the following data-generating model:
\begin{equation} \label{Model:FP}
    Y = 0.1 + x_1^{0.5} + x_1 + x_3^{-0.5}+ x_3^{-0.5} \log(x_3) + x_{4a} + x_5^{-1} + \log(x_6) + x_8 + x_{10} + \varepsilon, \varepsilon \sim N(0,1).
\end{equation}
This is similar to a model studied by \citet{royston2008multivariable}, but it is rendered more challenging by incorporating more nonlinearities than the original. 

To perform fractional polynomial analysis with the \pkg{FBMS} package, we specify the fractional polynomials of order 1 and 2 as transforms and ensure that the \code{gmjmcmc} algorithm uses only modifications to generate new features. This is achieved by setting \code{probs$gen <- c(0,1,0,1)}. Furthermore, we set the maximum feature depth to $D = 1$, which guarantees that only fractional polynomials of order 1 and 2 are potentially included in the model.

To this end, we first generate a \code{params} list with default values using \code{gen.params.gmjmcmc()}, which requires the number of input covariates as an argument (in this case \code{ncol(df) - 1}). The maximum depth is then specified via \code{params$feat$D}. We would like to once again emphasize that a comprehensive description of all parameters in the \code{params} list is provided in Appendix~\ref{subset:gen.params}.

\begin{CodeChunk}
\begin{CodeInput}
R> transforms <- c("p0","p2","p3","p05","pm05","pm1","pm2",
                   "p0p0","p0p05","p0p1","p0p2","p0p3",
                   "p0p05","p0pm05","p0pm1","p0pm2")
R> probs <- gen.probs.gmjmcmc(transforms)
R> probs$gen <- c(0,1,0,1) # Only modifications!
R> params <- gen.params.gmjmcmc(ncol(df) - 1)
R> params$feat$D <- 1   # Set depth of features to 1
\end{CodeInput}
\end{CodeChunk}

The analysis itself is then performed as before, where we start again with the single thread analysis with default settings.
            
\begin{CodeChunk}
\begin{CodeInput}
R> result <- fbms(data = df, method = "gmjmcmc", transforms = transforms, 
R>                 probs = probs, params = params)
R> summary(result)
\end{CodeInput}
\end{CodeChunk}

\begin{CodeChunk}
\begin{CodeOutput}
  feats.strings   marg.probs
1        p0(x6) 1.0000000000
2           x10 1.0000000000
3            x1 1.0000000000
4            x8 0.9989068571
5            x6 0.4782058635
6      p0p3(x6) 0.0003752229
\end{CodeOutput}
\end{CodeChunk}

As expected these results are not entirely convincing. The linear terms $x_1$, $x_8$, $x_{10}$ as well as the nonlinear term $\log(x_6)$ are correctly included. However, there are additional terms involving  $x_6$, and other linear and nonlinear features from the data generating model including the terms $x_1$, $x_3$, $x_{4a}$ and $x_5$ are missing. Also running the single-thread default analysis with different seeds yields entirely different models.
We next use one more time \code{gmjmcmc.parallel} with forty parallel runs.

\begin{CodeChunk}
\begin{CodeInput}
R> result_parallel <- fbms(data = df, transforms = transforms, 
R>                         probs = probs, params = params, P=25,
R>                         method = "gmjmcmc.parallel",runs=40, cores=40)
R> summary(result_parallel, tol = 0.01)
\end{CodeInput}
\end{CodeChunk}

\begin{CodeChunk}
\begin{CodeOutput}
  feats.strings marg.probs
1           x10 1.00000000
2            x1 0.99999994
3        p0(x6) 0.99996160
4            x8 0.99955073
5           x4a 0.91556748
6            x6 0.02621301
7            x5 0.01363549
8           x4b 0.01345450
\end{CodeOutput}
\end{CodeChunk}

The multiple thread analysis with a relatively small number of MJMCMC iterations within the populations gives already quite a good model, only missing $\sqrt{x_1}$, $x_5^{-1}$, and the nonlinear features  involving $x_3$. Increasing the number of MJMCMC iterations in each thread yields a fairly similar model with correctly identified terms only.

\begin{CodeChunk}
\begin{CodeInput}
R> result_parallel2 <- fbms(data = df, transforms = transforms, 
R>                         probs = probs, params = params, P=25, 
R>                         N=1000, N.final=2000, 
R>                         method = "gmjmcmc.parallel", runs=40, cores=40)
R> summary(result_parallel2, tol = 0.05)
\end{CodeInput}
\end{CodeChunk}

\begin{CodeChunk}
\begin{CodeOutput}
   feats.strings marg.probs
1            x10 1.00000000
2             x1 0.99997031
3         p0(x6) 0.99995748
4             x8 0.99892919
5            x4a 0.93667332
\end{CodeOutput}
\end{CodeChunk}
Now, all the linear terms are correctly specified and also $\log(x_6)$ is correct. However, nonlinear features  involving $x_3$ and $x_5$ are still missing.

Increasing the number of populations and MJMCMC iterations does not really change the model, which indicates that the Markov chain has more or less converged in this example. This also seems to be confirmed when inspecting the diagnostic plots. Changing the coefficient prior can  improve the results, for example with Jeffreys-BIC prior some nonlinear terms involving $x_5$ tend to be detected. More on the choice of priors will be discussed in Section \ref{Sec:Priors}.


\subsection[BGNLM]{Example 5: nonlinear Projections}\label{SubSec:BGNLM}

In the following example, we focus on the predictive ability of models that include only nonlinear projections. To this end, we study the abalone data set, which was first introduced in \cite{waugh1995extending}. The data set is available in our package via \code{data(abalone)}.

The task is to predict the age of abalone shells based on several physical measurements, which are easier to obtain than counting the number of rings that actually determine age. The available predictors include \textit{sex} (a categorical variable with three levels for abalones), as well as the continuous variables \textit{length, diameter, height, whole weight, shucked weight, viscera weight}, and \textit{shell weight}.
This prediction task is particularly challenging due to strong collinearity among the predictors.

After importing the data, we create two dummy variables to represent the three categories of \textit{sex}, using the category \textit{Infant} as the baseline.
To assess the predictive performance of different modeling strategies, we split the data into a training and a test set, following the same partition used in \citet{waugh1995extending}.
Our primary analysis will focus on models that use only projections based on the \code{sigmoid} function as the nonlinear transformation. This approach closely resembles a classical neural network, although our algorithm still permits the inclusion of original covariates in the model.

\begin{CodeChunk}
\begin{CodeInput}
R> transforms <- c("sigmoid")
R> probs <- gen.probs.gmjmcmc(transforms)
R> probs$gen <- c(0,0,1,1) #Only projections!
\end{CodeInput}
\end{CodeChunk}

We first consider again the analysis using only one thread.

\begin{CodeChunk}
\begin{CodeInput}
R> result = fbms(data = df.training, method = "gmjmcmc",
                 transforms = transforms, probs = probs)
R> summary(result)
\end{CodeInput}
\end{CodeChunk}

\begin{CodeChunk}
\begin{CodeOutput}
                     feats.strings  marg.probs
1                         Diameter 1.000000000
2                           Height 1.000000000
3                         Weight_W 1.000000000
4                         Weight_S 1.000000000
5                         Weight_V 1.000000000
6                       Sex_F_vs_I 1.000000000
7                       Sex_M_vs_I 1.000000000
8                           Length 1.000000000
9     sigmoid(1+1*Length+1*Height) 1.000000000
10                       Weight_Sh 0.999294909
11 sigmoid(1+1*Weight_Sh+1*Height) 0.009856509\end{CodeOutput}
\end{CodeChunk}

Note that all variables have entered the final model as linear predictors. In addition, there is one nonlinear feature of depth one. We now evaluate how the trained model performs on the test data set:

\begin{CodeChunk}
\begin{CodeInput}
R> pred = predict(result, x =  df.test[,-1])  
R> sqrt(mean((pred$aggr$mean - df.test$Rings)^2))
R> plot(pred$aggr$mean, df.test$Rings)
\end{CodeInput}
\end{CodeChunk}
\begin{figure}[ht!] 
    \centering
\includegraphics[width=0.9\textwidth]{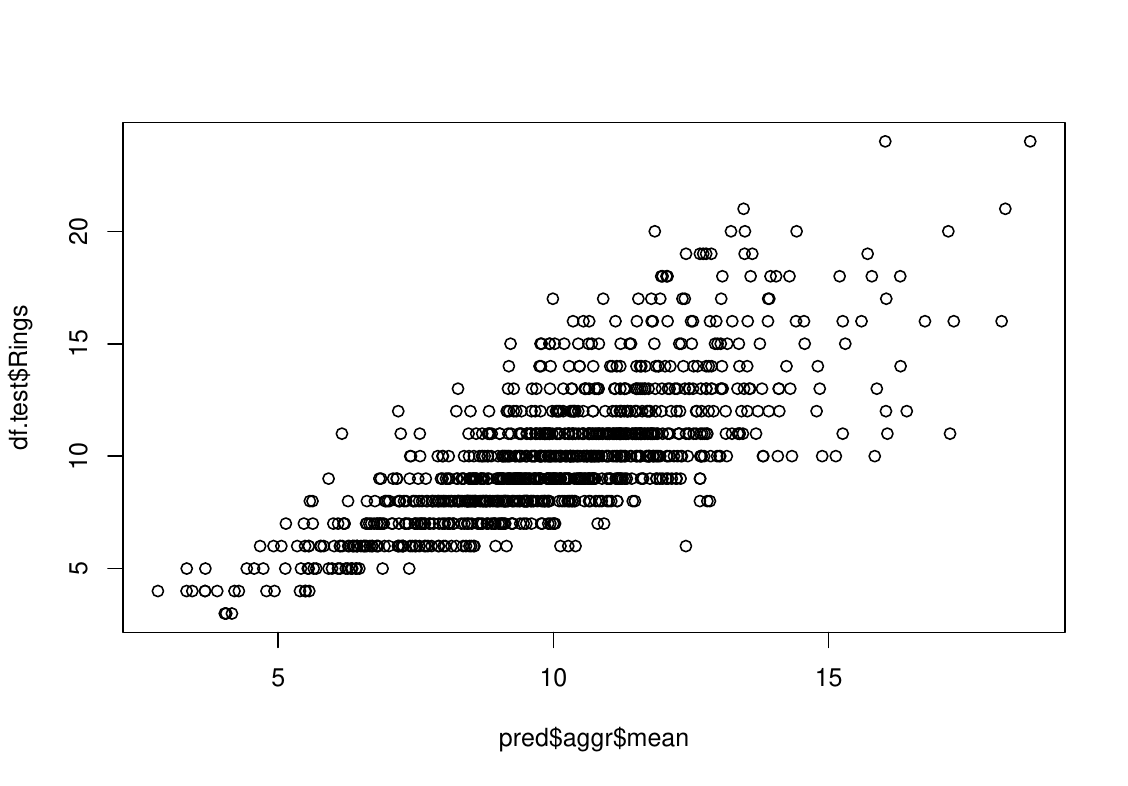}
\end{figure}
The root mean of the prediction residual sum of squares (RMSE) is approximately 2.078 and the prediction plot illustrates that specifically for older snails the prediction model is not very precise. This coincides with the findings from previous studies on this data set. 

We once again want to compare the performance with a model which was built using 40 parallel threads.

\begin{CodeChunk}
\begin{CodeInput}
R> result_parallel = fbms(data = df.training, method = "gmjmcmc.parallel",
R>                        transforms = transforms, probs = probs, P=25,
R>                         runs = 40, cores = 40, )
R> summary(result_parallel)
\end{CodeInput}
\end{CodeChunk}

\begin{CodeChunk}
\begin{CodeOutput}
             feats.strings marg.probs
1               Weight_Sh 1.00000000
2                Weight_V 1.00000000
3   sigmoid(1+1*Weight_S) 1.00000000
4                Weight_W 0.99999999
5              Sex_M_vs_I 0.99998851
6              Sex_F_vs_I 0.99997317
7                  Height 0.99985312
8                  Length 0.99958332
9                Diameter 0.99935956
10    sigmoid(1+1*Length) 0.96958318
11               Weight_S 0.91567421
12  sigmoid(1+1*Weight_W) 0.88525741
13 sigmoid(1+1*Weight_Sh) 0.08432579
\end{CodeOutput}
\end{CodeChunk}
Again, all the initial covariates are included in the model, but this time four additional nonlinear features, each with depth $d = 1$, are also selected. The prediction root mean squared error on the test data set is approximately 2.065, which is slightly lower than that of the single-threaded model.

The reader may have noticed that the $\alpha$  coefficients for the nonlinear projections are all equal to 1. This is because, by default, the simplest strategy is used, in which the $\alpha$ coefficients are not estimated but are instead fixed at 1. While this method is computationally efficient, it is not necessarily the most desirable in terms of modeling flexibility or predictive accuracy. The \pkg{FBMS} package currently implements two alternative strategies for estimating the $\alpha$  coefficients. One is called \code{"deep"}, corresponding to the third method described by \citet{hubin2021flexible}. The other one named \code{"random"}, adopts a fully Bayesian approach by placing standard normal priors on the internal parameters of the features.

To use the second method for estimating the $\alpha$ coefficients, we need to set
\begin{CodeChunk}
\begin{CodeInput}
R> params$feat$alpha = "deep"
\end{CodeInput}
\end{CodeChunk}
Subsequently, the same code as before can be used to obtain results for both the single-threaded and parallel analyses. For comparison, we also conducted a fractional polynomial analysis. Table~\ref{tab:res_abalone} reports the root mean squared errors for all five methods. Overall, the best predictive performance was achieved using \code{gmjmcmc.parallel} with nonlinear projections and the \code{"deep"} method for estimating $\alpha$. Interestingly, in the single-threaded analysis, fixing the coefficients to 1 resulted in a lower RMSE than the \code{"deep"} method. However, the differences in predictive performance across methods are generally small.

\begin{table}[htb!]
\centering
\begin{tabular}{lccccc}  
\hline
   & $\alpha = 1$ & $\alpha = 1$ & $\alpha$ deep & $\alpha$ deep &  Frac. pol.\\ 
  &  single thread & parallel & single thread & parallel &  parallel \\ 
  \hline
RMSE & 2.078 & 2.065 & 2.101 & 2.035 & 2.072 \\ 
   \hline
\end{tabular}
\caption{\label{tab:res_abalone}
Prediction root mean squared error for five different modeling strategies, where predictions were performed based on model avaraging over the best population.}
\end{table}


\section[Priors]{Specifying Priors}\label{Sec:Priors}

Having learned how to generate specific (nonlinear) features within a Gaussian model framework for metric outcomes, we now illustrate how to modify priors within the \pkg{FBMS} package. The example in Section~\ref{SubSec:GPrior} considers high-dimensional data analyzed with linear models after preselecting candidate predictors. In this setting, the hyper-parameters of the priors on the model coefficients need to be adjusted. This also allows us to learn how to specify custom hyper-parameter values in the \pkg{FBMS} package.

In the example from Section~\ref{SubSec:model_prior} we aim to use an alternative complexity measure for the model prior $p(\M)$. The key to implementing such custom priors is to define a corresponding function, which is then passed to the \code{fbms} function via the \code{loglik.pi} argument. Section~\ref{SubSec:model_prior} provides a comprehensive introduction to the syntax of this function, which will also be essential in Section~\ref{Sec:Extensions}, where we consider non-Gaussian models.

\subsection[G - Prior]{Example 6: Specification of hyper-parameter in a g-prior} \label{SubSec:GPrior}

To demonstrate how the \pkg{FBMS} package can be used to fit purely linear models in a high-dimensional setting, we introduce a classical dataset from human genetics. \citet{stranger2007} collected gene expression levels from lymphoblastoid cell lines of $n = 210$  unrelated individuals. The original dataset is available at {\it ftp://ftp.sanger.ac.uk/pub/genevar/}.

Our goal is to identify genes that regulate the expression level of the gene CCT8, which lies within the Down syndrome critical region on human chromosome 21. We treat the expression level of CCT8 as the quantitative response and the expression levels of all other genes as predictors. The full dataset includes expression measurements for $p = 47293$ probes.
Following \citet[][Section 1.6.1]{bogdan2020handbook}, we apply a preliminary filtering step by excluding probes where the difference between maximum and minimum expression levels is smaller than 2, resulting in a reduced set of $p = 3220$ predictors. This filtered dataset is included in the package under the name \code{SangerData2}.

The following code loads the dataset and shortens the column names for more concise output:
\begin{CodeChunk}
\begin{CodeInput}
R> data(SangerData2)
R> df = SangerData2)
R> colnames(df) = c("y",paste0("x",1:(ncol(df)-1)))
R> n = dim(df)[1]; p = dim(df)[2]-1
\end{CodeInput}
\end{CodeChunk}

Restricting the class of features $\mathcal{F}$ to include only linear effects can be accomplished by limiting the set of feature-generating operators to mutation alone, thus allowing only changes to the set of variables in the feature population. Specifically, setting \code{probs$gen = c(0, 0, 0, 1)} ensures that mutation is used with probability 1, while all other operators are disabled.

\begin{CodeChunk}
\begin{CodeInput}
R> transforms = c("")
R> probs = gen.probs.gmjmcmc(transforms)
R> probs$gen = c(0,0,0,1)
\end{CodeInput}
\end{CodeChunk}

By default, \code{fbms} includes all covariates in the initial population. However, when the number of covariates is large, it becomes computationally impractical to explore the full set simultaneously using the MJMCMC algorithm directly. To address this, we initialize the algorithm with a subset of promising candidate genes as predictors. Specifically, we preselect the 50 covariates that show the highest correlation with the response variable. This is implemented by assigning the corresponding index vector \code{ids} to the parameter \code{params$feat$prel.filter}, which defines the covariates used in the first population.

\begin{CodeChunk}
\begin{CodeInput}
R> c.vec = unlist(mclapply(2:ncol(df), function(x)abs(cor(df[,1],df[,x]))))
R> ids = sort(order(c.vec,decreasing=TRUE)[1:50]) 
R> params = gen.params.gmjmcmc(p)
R> params$feat$prel.select <- ids
\end{CodeInput}
\end{CodeChunk}
Additionally, to ensure that populations always include the same number of covariates, we set the population size equal to the number of preselected features so that both the initial and subsequent populations have the same size. 

\begin{CodeChunk}
\begin{CodeInput}
R> params$feat$pop.max <- 50    # Maximum population size 
\end{CodeInput}
\end{CodeChunk}
Obtaining reliable results for this large-scale problem requires both a sufficiently large number of MJMCMC iterations within each population and a sufficiently high number of populations. 
In this example, we report results based on $P = 50$ populations, each undergoing  $N = 1000$ MJMCMC iterations, and executed across 10 parallel runs.

Here, we must explicitly specify the hyperparameter of the g-prior based on the true number of covariates $p$, rather than the default choice used by the \code{fbms} function, which is the cardinality of the \code{ids} set of preselected covariates. 
This is achieved by including the argument \code{beta_prior} when calling \code{fbms}, which allows selecting any of the priors for the $\beta$ coefficients listed in Table \ref{tab:priors}.

\begin{CodeChunk}
\begin{CodeInput}
R> result_parallel1 = fbms(data=df,transforms=transforms, 
R>                    beta_prior = list(type="g-prior", g=max(n,p^2)),
R>                    method="gmjmcmc.parallel", probs=probs,params=params,
R>                    P=50,N=1000,runs=10,cores=10)
\end{CodeInput}
\end{CodeChunk}

\begin{figure}[t]
\centering
\begin{tabular}{cc}
\begin{tabular}{lrllll}
  \hline
\multicolumn{2}{c}{Run 1} & \multicolumn{2}{c}{Run 2} & \multicolumn{2}{c}{Run 3} \\
feats & prob & feats & prob & feats & prob \\ 
  \hline
  x1016 & 1.000 & x1016 & 1.000 & x1016 & 1 \\ 
  x1402 & 0.712 & x1402 & 0.29 & x1402 & 0.944 \\ 
  x2202 & 0.211 & x3042 & 0.226 & x1669 & 0.495 \\ 
  x183 & 0.157 & x1156 & 0.195 & x1658 & 0.455 \\ 
   &  & x1658 & 0.125 & x2202 & 0.105 \\ 
   &  & x3056 & 0.116 &  &  \\ 
   &  & x2202 & 0.107 &  &  \\  
   \hline
\end{tabular}&
\raisebox{-3cm}{\includegraphics[width=0.5\textwidth,trim={0 0cm 0 0},clip]{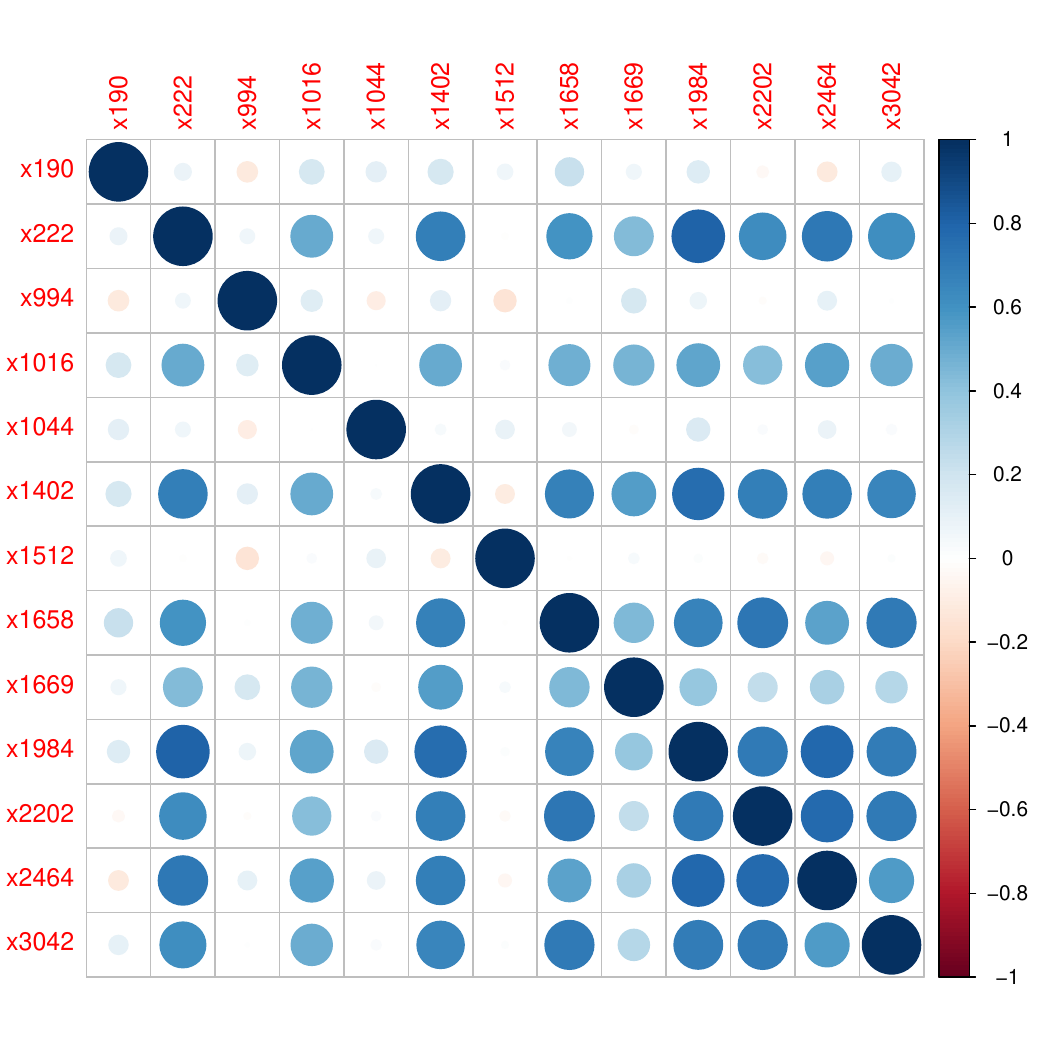}
}
\end{tabular}
\caption{\label{fig:Sanger.corr}Left: Selected features (listed in the feats column) and their corresponding estimated marginal posterior probabilities (in the prob column) for three different runs. Only features with marginal probabilities greater than 0.05 are shown. Results are based on the population with the highest model probability. Right: Pairwise correlations between covariates with (estimated) marginal posterior probabilities greater than 0.05 in at least one of the three parallel runs of the Sanger example analysis.}
\end{figure}

One such call to \texttt{gmjmcmc.parallel} took about 5 minutes to complete on an Intel(R) Xeon(R) Gold 6226R CPU @ 2.90GHz with 64 cores, two threads per core, and 772GB of RAM. Repeating these calls with different random seeds produced the results shown in Figure~\ref{fig:Sanger.corr}, where the first two columns correspond to the first run, the next two to the second run, and the last two to the third run. Only features with (estimated) marginal probabilities greater than 0.01 are included.

We observe that features $x1016$ and $x1402$ consistently have the highest marginal probabilities across all three runs. None of the other features appear in the median probability model for any run. Figure~\ref{fig:Sanger.corr}, right panel, presents the pairwise correlations among all covariates listed in the left part of Figure~\ref{fig:Sanger.corr}. The presence of very high correlations among many variables highlights the difficulty of this problem.

%


\subsection[prior]{Example 7: Change of model prior with custom function}\label{SubSec:model_prior}

The GMJMCMC algorithm was originally introduced by \citet{hubin2018novel} in the context of Bayesian logic regression, where logical expressions of binary covariates coded as 0 (FALSE) and 1 (TRUE) serve as predictor variables. We now illustrate how such models can be implemented using the \pkg{FBMS} package, focusing again on a metric outcome with Gaussian error terms.

To generate Boolean features, we define the three basic logical operations: AND, OR, and NOT. As shown in Table~\ref{Tab:functions}, the logical complement $x^c$ is implemented in the \pkg{FBMS} package using the function \code{not}, which is simply defined as not$(x) = 1-x$. The logical AND between two binary variables corresponds to the product $x_1\wedge x_2 = x_1*x_2$. The logical OR is implemented using De Morgan's law
$$
x_1 \vee x_2 = \mbox{not}(\mbox{not}(x_1)*\mbox{not}(x_2))
$$
Thus, to represent all three logical operations in \pkg{FBMS}, we require only the \code{not} transformation in modifications and multiplicative interactions for logical AND. No projection operators are needed. Therefore, we will set
\begin{CodeChunk}
\begin{CodeInput}
R> transforms = c("not")
R> probs$gen <- c(1, 1, 0, 1)
\end{CodeInput}
\end{CodeChunk}

We follow the approach of \citet{hubin2018novel} in specifying priors, using Jeffreys prior for the parameters in combination with a model prior that penalizes each feature $F_j(\bm x)$ according to the number of its leaves $s(F_j(\bm x))$ which corresponds to the total width $w_j$ of the feature \citet[ see for detail on the total width definition]{hubin2021flexible}. 
The multiplicative contribution of a specific tree of size $s$ to the model prior is inversely proportional to the total number of trees $N(s)$  of the same size  $s$.  

Instead of the model prior (\ref{glmgammaprior}) we thus consider 
\begin{equation}\label{Prior.Model.Logic}  
p(\M) \propto \prod_j\   \left(N(s(F_j(\bm x)))\right)^{-\gamma_j}, \quad \mbox{with } s(F_j(\bm x)) \leq C_{max},   
\end{equation}
where $\gamma_j$ denotes the trees included in the model and $C_{max}$ denotes an upper bound on the total number of leaves which are allowed in a logical tree. For example, the feature $\mbox{not}(\mbox{not}(x_1)*\mbox{not}(x_2))$  has two leaves, just like the feature $x_1*x_2$. Thus, the complexity measure no longer counts the number of operations used to generate a feature, as was the default in previous examples. Instead, the penalty is now based on $s(F_j(\bm x))$, which corresponds to the total width $w_j$ of the feature $F_j(\bm x)$.

This penalty structure closely resembles a Bonferroni correction in multiple testing. Since the number of such trees, $N(s)$ , grows exponentially with tree size $s$, this choice naturally assigns smaller prior probabilities to larger trees. However, computing $N(s)$ exactly becomes computationally infeasible for large $s$. To obtain a rough approximation, \citet{hubin2018novel} ignored logical expressions involving repeated use of the same variable and derived the following estimate for the number of trees of width $w_j$:
$$
\hat{N}(w_j) \approx \binom{p}{w_j}\ 2^{2w_j-2} \approx \frac{(4p)^{w_j}}{4 w_j!}.
$$

\subsubsection{Function to compute the log posterior}

Having defined the parameter and model priors, we now demonstrate how to implement them in practice. This is the first example that illustrates the use of custom functions for computing the marginal model posterior within the \pkg{FBMS} package, which will also play an important role in Section \ref{Sec:Extensions}.
To enable this, the function to compute the log posterior must be defined in a specific format, as required by the package:
\begin{center}
    \code{function (y, x, model, complex, mlpost_params)} 
\end{center}
The arguments of the function, all of which are defined in the local environment of \code{gmjmcmc}, are described as follows:
\begin{itemize}
\item \code{y}: The response variable of the model.
\item \code{x}: The design matrix consisting of all features currently included in the population (not to be confused with input covariates). The first column is a vector of ones representing the intercept, except when this is explicitly excluded from the model.
\item \code{model}: A logical vector indicating which columns of \code{x} are included in the current model. If the intercept is part of the model, the first element of \code{model} will always be \code{TRUE}. Likewise, if the argument \code{fixed} specifies that certain covariates must be included, the corresponding elements of \code{model} will also be \code{TRUE}.
\item \code{complex}: A named list containing three elements that describe the complexity of each feature:
\begin{itemize}
\item \code{$oc} – operation count,
\item \code{$width} – total width,
\item \code{$depth} – depth of the feature.
\end{itemize}
These quantities are computed internally and can be used to define different model priors. For this example, we will use \code{$width} to implement the model prior (\ref{Prior.Model.Logic}) for logic regression.
\item \code{mlpost_params}: A list containing all parameters passed to \code{gmjmcmc} via the \code{model_prior} and \code{extra_params} arguments. It allows additional parameters to be supplied to custom functions and is used in this example to pass the total number of covariates, denoted by $p$. We will see this argument used again in Examples 9–12 in Section~\ref{Sec:Extensions}.
\end{itemize}
The function returns the sum of log prior and log marginal likelihood, i.e. up to a constant a log-posterior $\log p(\M|Y)$, which is assigned to \code{crit}, as well as the estimated coefficients from the linear model. These coefficients are subsequently used by the \code{predict} function.
Below is the function \code{estimate.logic.lm}, which calculates (up to an additive constant) the log-posterior for our example. It uses Jeffreys prior for the regression coefficients together with the model prior defined in equation~(\ref{Prior.Model.Logic}).
\begin{CodeChunk}
\begin{CodeInput}
R> estimate.logic.lm = function(y, x, model, complex, params)
R> { 
R>   # Computation of marginal log-likelihood using Jeffreys prior
R>   suppressWarnings({
R>     mod <- fastglm(as.matrix(x[, model]), y, family = gaussian())
R>   })
R>   mloglik <- -(mod$aic + (log(length(y))-2) * (mod$rank))/2 
R>   
R>   # Computation of log of model prior
R>   wj <- complex$width
R>   lp <- sum(log(factorial(wj))) - sum(wj*log(4*mlpost_params$p) - log(4))
R>   
R>   logpost <- mloglik + lp #log posterior up to a constant
R>   
R>   if(logpost==-Inf)
R>     logpost = -10000
R>   
R>   return(list(crit = logpost, coefs = mod$coefficients))
R> }
\end{CodeInput}
\end{CodeChunk}

The line \code{mloglik <- -(mod$aic + (log(length(y))-2) * (mod$rank))/2} computes the marginal likelihood
under Jeffreys prior with unknown variance of the response. The implementation is exactly like in the \code{gaussian.loglik} function, except for the model prior. \code{gaussian.loglik} uses the default log-prior which is coded as
\begin{CodeChunk}
\begin{CodeInput}
R> log(params$r) * (sum(complex$oc))
\end{CodeInput}
\end{CodeChunk}
using the operation count as complexity measure. In contrast we are using here the total width \code{wj <- complex$width}. This is defined as the sum of the local widths of all features involved in generating a new feature and in case of logic regression is nothing else but the total number of leaves of a tree. 

The expression
\code{sum(log(factorial(wj))) - sum(wj*log(4*mlpost_params$p) - log(4))}
computes the logarithm of the prior, as defined in Equation~(\ref{Prior.Model.Logic}). Here, \code{mlpost_params$p} represents the total number of possible leaves, which corresponds to the number of input covariates $p$. This value is passed to the function via the argument \code{model_prior = list(p=p)} in the \code{fbms} call, as described below.
Finally, the log-posterior (up to an additive constant) is computed as
\code{logpost <- mloglik + lp}.

We evaluate our implementation of Bayesian logic regression using simulation scenario 5 from \citet{hubin2018novel}, which features $p = 50$  randomly generated binary covariates and includes logical trees with 1 to 4 leaves in the data-generating process:
\begin{equation}\label{modeleq.logic}
  Y = 1  + 1.5 x_{37} + 3.5 (x_2 \wedge x_9) +
  9 (x_7 \wedge x_{12} \wedge x_{20}) +  
  7 (x_4 \wedge x_{10} \wedge x_{17} \wedge x_{30}) +\epsilon \; .
\end{equation} 
Data for $n = 2000$ samples are generated and evenly split into training and test sets. For the test set, the mean response values from the data-generating model are recorded.

To perform data analysis with \code{fbms}, we set \code{probs} as described above and configure the \code{params} argument to restrict the maximum population size to 50 and the maximum number of leaves per tree to 15.
\begin{CodeChunk}
\begin{CodeInput}
R> params$feat$pop.max <- 50
R> params$feat$L <- 15   #C_max
\end{CodeInput}
\end{CodeChunk}
Single and parallel GMJMCMC are then run as follows
\begin{CodeChunk}
\begin{CodeInput}
R> result <- fbms(formula = Y2~1+., data = df.training, 
R>           probs = probs, params = params,  
R>           method = "gmjmcmc", transforms = transforms, N = 500, P = 25,
R>           family = "custom", loglik.pi = estimate.logic.lm,
R>           model_prior = list(p = p))
R>
R> result_parallel <-  fbms(formula = Y2~1+.,data = df.training, 
R>           probs = probs, params = params, 
R>           method = "gmjmcmc.parallel", transforms = transforms, 
R>           N = 500, P=25, runs = 16, cores = 8, 
R>           family = "custom", loglik.pi = estimate.logic.lm, 
R>           model_prior = list(p = p))
\end{CodeInput}
\end{CodeChunk}

By now, we are familiar with most of the arguments in the \code{fbms} function used above, except for the last two lines in each call. The argument \code{family = "custom"}, combined with \code{loglik.pi = estimate.logic.lm}, invokes the function \code{estimate.logic.lm} to compute the log-posterior. This function requires the parameter \code{model_prior$p} to evaluate the posterior, as described in equation~(\ref{Prior.Model.Logic}). 

The required parameter is passed to \code{estimate.logic.lm} via the argument \code{model_prior = list(p = p)}. Alternatively, \code{extra_params = list(p = p)} could be used; however, our convention is to use \code{model_prior} for parameters needed to compute the model prior, \code{beta_prior} for parameters needed to compute the parameters' prior, and \code{extra_params} for parameters required to compute the marginal likelihood in a custom function or to provide additional inputs to the models. In Appendix~\ref{SubSec:fbms}, we illustrate how to implement the robust g-prior as a special case of a tCCH prior with the dispersion parameter fixed at 1, in which case several parameters are additionally passed via \code{beta_prior}.

Using Jeffreys prior, both the single-threaded and parallel versions of the analysis both the single-tread and the parallel versions get the same (correct) results. For the sake of brevity, we focus here on the results from the single-threaded version.
\begin{CodeChunk}
\begin{CodeInput}
R> summary(result)
\end{CodeInput}
\end{CodeChunk}

\begin{CodeChunk}
\begin{CodeOutput}
         feats.strings   marg.probs
1              (V9*V2) 1.0000000000
2 (((V4*V10)*V17)*V30) 1.0000000000
3                  V37 1.0000000000
4       ((V20*V7)*V12) 1.0000000000
5                  V20 0.0006767398
\end{CodeOutput}
\end{CodeChunk}

We observe that all trees from the data-generating model have a posterior probability of 1. Consequently, the median probability model coincides with the true data-generating model:
\begin{CodeChunk}
\begin{CodeInput}
R> mpm <- get.mpm.model(result, y = df.training$Y2, x = df.training[,-1])
R> mpm$coefs
\end{CodeInput}
\end{CodeChunk}
\begin{CodeChunk}
\begin{CodeOutput}
  (Intercept)   (V9*V2)  (((V4*V10)*V17)*V30)        V37   ((V20*V7)*V12)  
     1.113641  3.419107              6.894541   1.343824         9.055281
\end{CodeOutput}
\end{CodeChunk}
The model with the highest posterior probability is obtained using \code{get.best.model(result)} and, in this example, coincides with the median probability model (MPM).
Predictions can be made either using model averaging or based on the median probability or best models. Since nearly all of the posterior mass is concentrated on the best model, all three approaches yield identical results. Their predictive performance closely matches that of the mean from the true data-generating model, with root mean square errors of 
1.0344 and 1.0282, respectively.


\section[Extensions]{Beyond the Gaussian Model}\label{Sec:Extensions}
%

In this section, we illustrate how \pkg{FBMS} can be applied to a broad class of statistical models. Direct implementations of marginal posterior probability computations are currently available for the Gaussian model and for generalized linear models (GLMs) with the "binomial", "poisson", and "gamma" families. Table~\ref{tab:priors} lists all priors currently supported for GLMs. In Section~\ref{SubSec:GAM}, we demonstrate how to fit a simple logistic regression model for a binary outcome variable.

However, by using the \code{loglik.pi} argument, it is possible to specify virtually any model for which the marginal posterior probability can be computed within a reasonable amount of time. This section demonstrates how to implement more of such custom models, including examples involving a linear mixed model, a Poisson mixed model, a subsampling-based approach for handling tall data, and Cox regression for survival analysis.


\subsection[Binary]{Example 8: Binary Response} \label{SubSec:GAM}

Let's consider the logistic regression model for a binary response, that is $Y_i,  i \in \{1,\dots,n\} \in \{0, 1\}$: 

\begin{align}\label{logistic.regression}
   P(Y_i = 1)&=\pi_i \nonumber\\
 \mbox{logit}(\pi_i)&=\beta_0  + \sum_{j=1}^{q} \gamma_{j}\beta_{j}F_{j}(\bm{x}_i).
\end{align}

The corresponding likelihood, combined with Jeffreys prior, can be specified in the \pkg{FBMS} package using \code{family = "binomial"} and \code{beta_prior = list(type = "Jeffreys-BIC")}. We illustrate this setup by analyzing the spam dataset from \citet{cranor1998spam}, which is included in the \pkg{kernlab} package.
It contains a collection of both spam and non-spam emails, with a total of 4,601 emails, of which 1,813 are labeled as spam.

For each email, 57 continuous variables are provided and can serve as predictors in a classification model. Although we could split the data into training and test sets, our primary goal here is to demonstrate how to perform logistic regression analysis. Therefore, we keep the setup as simple as possible. We apply the same set of transformations as in the introductory example in Section~\ref{Sec:BasicIntro}, with the exception that we disable the checks for multicollinearity as the main goal here is prediction:

\begin{CodeChunk}
\begin{CodeInput}
R> params$feat$check.col <- F
\end{CodeInput}
\end{CodeChunk}

We can then call \code{fbms} to run GMJMCMC with default settings. 
\begin{CodeChunk}
\begin{CodeInput}
R> result <- fbms(formula = y~1+.,data = df, method = "gmjmcmc", 
R>               family = "binomial", beta_prior = list(type = "Jeffreys-BIC"),
R>               transforms = transforms, probs = probs, params = params)
\end{CodeInput}
\end{CodeChunk}
The output obtained using \code{summary} includes five nonlinear features with posterior probabilities greater than 0.9.
\begin{CodeChunk}
\begin{CodeOutput}
   feats.strings   marg.probs
1             x5 1.0000000000
2             x8 1.0000000000
3        p0(x57) 1.0000000000
4            x16 1.0000000000
5            x21 1.0000000000
6            x24 1.0000000000
7            x25 1.0000000000
8            x27 1.0000000000
9     troot(x53) 1.0000000000
10           x29 1.0000000000
11           x33 1.0000000000
12  sigmoid(x23) 1.0000000000
13           x41 1.0000000000
14           x42 1.0000000000
15           x44 1.0000000000
16           x45 1.0000000000
17           x46 1.0000000000
18           x52 1.0000000000
19            x7 1.0000000000
20           x17 1.0000000000
21            x6 1.0000000000
22      (x11*x4) 1.0000000000
23      to3(x49) 0.9997947929
24           x20 0.9994774881
25           x22 0.9913972530
\end{CodeOutput}
\end{CodeChunk}
Next, for the sake of giving an example, we compute the in-sample classification accuracy. Care must be taken when using the model to make predictions, as the inverse link function must be explicitly specified when calling \code{pred}.
\begin{CodeChunk}
\begin{CodeInput}
R> pred = predict(result, x =  df[,-1], link = function(x)(1/(1+exp(-x))))  
R> mean(round(pred$aggr$mean)==df$y)
\end{CodeInput}
\end{CodeChunk}
The single-threaded analysis yields a classification accuracy of 93.9\%. Although the parallel version tends to include slightly more nonlinear terms, it results in only a modest improvement in terms of classification, with an accuracy of 94.3\%.


\newpage

\subsection[MM]{Example 9: Linear Mixed Model}\label{SubSec:MixedModel}
In this example, we use three different methods to compute the marginal log-likelihood for a linear mixed model, applied to a dataset on child undernutrition across regions of Zambia. This dataset was previously used by \citet{safken2021conditional} to evaluate their variable selection approach based on the conditional AIC criterion (cAIC). It is available through the \pkg{cAIC4} package and has been analyzed in several previous studies.

The outcome is a standardized measure of a child’s height. The dataset includes two potential random effects: the region (\code{reg}) and the district within the region (\code{dr}). Based on the cAIC criterion, a random intercept for \code{dr} was found to be optimal, and we will use only this random effect in our model. For feature generation, we consider the following four covariates: duration of breastfeeding (\code{c.bf}), age of the child (\code{c.age}), as well as the mother’s height (\code{m.ht}) and body mass index (\code{m.bmi}).

We begin by loading the data and scaling both the outcome and the predictor variables. 
\begin{CodeChunk}
\begin{CodeInput}
R> data(Zambia, package = "cAIC4")
R> df <- as.data.frame(sapply(Zambia[1:5],scale))
\end{CodeInput}
\end{CodeChunk}
Note that the district variable (\code{dr}), which will serve as a random effect, has not been included in \code{df}. Instead, it is passed via the \code{extra_params} list.

Next, we define the transformations and set the parameters for fractional polynomials, as described in Section~\ref{SubSec:FracPol}, but now we also allow for interaction terms.
\begin{CodeChunk}
\begin{CodeInput}
R> transforms <- c("p0","p2","p3","p05","pm05","pm1","pm2","p0p0","p0p05",
R>                 "p0p1","p0p2","p0p3","p0p05","p0pm05","p0pm1","p0pm2")
R> probs <- gen.probs.gmjmcmc(transforms)
R> probs$gen <- c(1,1,0,1) # Modifications and interactions!
R>
R> params <- gen.params.gmjmcmc(df)
R> params$feat$D <- 1  # Set depth to 1 (still allows interactions)
R> params$feat$pop.max = 10
\end{CodeInput}
\end{CodeChunk}
Since there are only four predictors, we keep the maximum population size relatively small by setting \code{params$feat$pop.max = 10}. As in Example 4 on fractional polynomials, we also set the feature depth to 1. This choice still allows interactions between covariates to be included in the model.

\subsubsection{lme4}
Our first approach uses the \pkg{lme4} package in combination with the Laplace approximation to compute marginal posteriors. As in Example 7, we need to define a custom function, here called \code{mixed.model.loglik.lme4}, to compute the logarithm of the likelihood evaluated at the posterior mode.

\begin{CodeChunk}
\begin{CodeInput}
R> mixed.model.loglik.lme4 <- function (y, x, model, complex, mlpost_params) 
R> { 
R>   # logarithm of marginal likelihood (Laplace approximation)
R>   if (sum(model) > 1) {
R>     x.model = x[,model]
R>     data <- data.frame(y, x = x.model[,-1], dr = mlpost_params$dr)
R>     
R>     mm <- lmer(as.formula(paste0("y ~ 1 +",
R>                paste0(names(data)[2:(dim(data)[2]-1)],
R>                collapse = "+"), "+ (1 | dr)")), data = data, REML = FALSE)
R>   } else{   #model without fixed effects
R>     data <- data.frame(y, dr = mlpost_params$dr)
R>     mm <- lmer(as.formula(paste0("y ~ 1 + (1 | dr)")), data=data, REML=F)
R>   }
R>   #Laplace approximation for beta prior
R>   mloglik <- as.numeric(logLik(mm)) - 0.5*log(length(y)) * (dim(data)[2]-2) 
R>   
R>   # logarithm of model prior
R>   if (length(mlpost_params$r) == 0)  mlpost_params$r <- 1/dim(x)[1]  
R>   lp <- log_prior(mlpost_params, complex)
R>  
R>   return(list(crit = mloglik + lp, coefs = fixef(mm)))
R> }
\end{CodeInput}
\end{CodeChunk}
The computations are fairly straightforward. The \code{lmer} function is used to fit the mixed model, with \code{REML = FALSE} specified to obtain maximum likelihood estimates. When constructing the model formula, it is important to distinguish between cases with and without fixed effects. Special care is needed when writing the formulas, since both \code{x} and \code{model} include the intercept. For this reason, only \code{x.model[, -1]} is included in the \code{data} dataframe. Consequently, \code{names(data)[2:(dim(data)[2] - 1)]} will contain exactly the names of the fixed effects included in \code{model}.

\subsubsection{INLA and RTMB}

Using the maximum likelihood estimate in combination with the Laplace approximation provides a simple and efficient way to approximate the marginal likelihood. To explore approaches that can be extended to more complex models, we consider two additional methods. First, we use \pkg{INLA} \citep{bakka2018spatial}, which is well known in the Bayesian community as an alternative to MCMC methods for approximate Bayesian inference. Second, we consider the \pkg{RTMB} package \citep{RTMB}, which provides an efficient implementation of the Laplace approximation, leveraging exact derivatives to compute marginal likelihoods.

While these packages offer an impressive range of modeling capabilities, their flexibility comes at the cost of increased computational time. To illustrate this, we analyze the Zambia dataset using \code{gmjmcmc}, generating only $P = 3$ populations and limiting the number of MJMCMC iterations per population to 30. This configuration is not intended to produce meaningful inference, but rather to facilitate a fair comparison of runtimes.
\begin{CodeChunk}
\begin{CodeInput}
R> #lme4 
R> result1a <- fbms(formula = z ~ 1+., data = df, transforms = transforms,
R>          method = "gmjmcmc", probs = probs, params = params, P=3, N = 30, 
R>          family = "custom", loglik.pi = mixed.model.loglik.lme4,
R>          model_prior = list(r = 1/dim(df)[1]),
R>          extra_params = list(dr = droplevels(Zambia$dr)))  
R> #INLA 
R> result1b <- fbms(formula = z ~ 1+., data = df, transforms = transforms,
R>          method = "gmjmcmc", probs = probs, params = params, P=3, N = 30, 
R>          family = "custom", loglik.pi = mixed.model.loglik.inla,
R>          model_prior = list(r = 1/dim(df)[1]),
R>          extra_params = list(dr = droplevels(Zambia$dr), 
R>                              INLA.num.threads = 10))
R> #RTMB 
R> result1c <- fbms(formula = z ~ 1+., data = df, transforms = transforms,
R>          method = "gmjmcmc", probs = probs, params = params, P=3, N = 30, 
R>          family = "custom", loglik.pi = mixed.model.loglik.rtmb,
R>          model_prior = list(r = 1/dim(df)[1]),
R>          extra_params = list(dr = droplevels(Zambia$dr), 
R>                              nr_dr = sum((table(Zambia$dr))>0)))

\end{CodeInput}
\end{CodeChunk}

Note that these calls of \code{fbms} differ only in the \code{loglik.pi} function and the parameters specified in the \code{extra_params} list. The list always includes the random factor \code{Zambia$dr}. For INLA, it additionally includes the parameter \code{INLA.num.threads}, which defines the number of parallel threads used internally by INLA. RTMB, on the other hand, explicitly requires the number of districts, specified via \code{nr_dr}.

The functions \code{mixed.model.loglik.inla} and \code{mixed.model.loglik.rtmb} are provided in Section~\ref{SubSec:fbms}. We do not include a full description of these two functions here, but instead focus on the resulting runtimes. With the specifications outlined above, \code{gmjmcmc} using the \code{lmer} function from the \pkg{lme4} package typically completes in under 6 seconds. In contrast, \pkg{INLA} requires between 2 and 4 minutes and \pkg{RTMB} between 4 and 7 minutes to complete the same task. While it may be possible to further optimize the log-posterior computation functions for these packages, a substantial increase in runtime should generally be expected when using them.

\subsubsection{Parallel analysis with lme4}
To perform a realistic analysis of the Zambia dataset, we use \code{gmjmcmc.parallel} with 40 parallel chains. We keep the maximum population size restricted at 10 and limit the number of MJMCMC iterations to 100. This choice is motivated by the fact that the dataset contains only four initial covariates, resulting in a relatively small number of possible feature combinations.
\begin{CodeChunk}
\begin{CodeInput}
R> result2a <- fbms(formula = z ~ 1+., data = df, transforms = transforms,
R>                  probs = probs, params = params, P=25, N = 100,
R>                  method = "gmjmcmc.parallel", runs = 40, cores = 40,
R>                  family = "custom", loglik.pi = mixed.model.loglik.lme4,
R>                  model_prior = list(r = 1/dim(df)[1]), 
R>                  extra_params = list(dr = droplevels(Zambia$dr)))
R>
R> summary(result2a,tol = 0.05,labels=names(df)[-1])   

\end{CodeInput}
\end{CodeChunk}

\begin{CodeChunk}
\begin{CodeOutput}
  feats.strings marg.probs
1         c.age  1.0000000
2          m.ht  0.9999997
3         m.bmi  0.9719209
4          c.bf  0.9709486
5  p0p05(c.age)  0.7415790
6 (c.age*c.age)  0.2582786
\end{CodeOutput}
\end{CodeChunk}
The variables \code{c.age}, \code{m.ht}, \code{ m.bmi}, \code{c.bf}, as well as some nonlinear features involving \code{c.age}, exhibit high posterior probabilities. We have rerun this example using 120 and 200 parallel chains, obtaining very similar results in each case, which suggests that the algorithm has converged.

It is interesting to compare our results with those of \citet{safken2021conditional}, who also suggest a nonlinear effect for \code{c.age} based on splines and include the other three variables as linear effects. Thus their findings are quite similar.


\subsection[MMP]{Example 10: Mixed-Effects Poisson Regression}\label{SubSec:MixedPoisson}

The following example highlights the capacity of the \pkg{FBMS} package to fit rather sophisticated models. Specifically, we employ \pkg{INLA} to compute the marginal log-likelihood for a Poisson regression model with random effects to account for repeated measurements, while using fractional polynomials to capture nonlinear relationships. The code for the corresponding \code{loglik.pi} function closely resembles that used in the previous linear mixed model with \pkg{INLA} and is again provided in Section~\ref{SubSec:fbms}. It is worth emphasizing that fitting such a complex nonlinear model while simultaneously performing fractional polynomial model selection would be extremely challenging using existing software solutions.

We illustrate how to fit this model using the \code{Epil} dataset, which is included in the \pkg{INLA} R package and briefly described in Volume I of the OpenBUGS Manual (\url{https://webbugs.psychstat.org/wiki/Manuals/Examples/Epil.html}). The dataset contains epileptic seizure counts for 59 patients, recorded over four consecutive time periods. Potential predictor variables include the baseline number of seizures (\code{Base}), the patient's age, and indicator variables for each visit. Our goal is to fit a Poisson regression model for the seizure counts, incorporating random intercepts for each patient to account for the repeated measurements structure.

We start with loading the data set 
\begin{CodeChunk}
\begin{CodeInput}
R> library(INLA)
R> data = INLA::Epil
R> df = data[1:5]
\end{CodeInput}
\end{CodeChunk}
As in the previous example, the parameters for fractional polynomial analysis with interactions are set up identically, and the patient-level random intercept data are again provided via the \code{extra_params} list.

The following settings adjust several tuning parameters of the MJMCMC process. These modifications are intended to speed up the algorithm, as computing the marginal likelihood using INLA is even more time-consuming for this model than for the linear mixed model. At the same time, this example demonstrates how to apply more advanced tuning of the MJMCMC algorithm. Specifically, we restrict the greedy local optimizers to only two iterations, each with a single proposal, and configure the simulated annealing optimizers to use a temperature step of 10 while cooling only down to 0.1. Further details on these tuning parameters can be found in Appendix \ref{subset:gen.params}.

\begin{CodeChunk}
\begin{CodeInput}
R> params$greedy$steps = 2
R> params$greedy$tries = 1
R> params$sa$t.min = 0.1
R> params$sa$dt = 10
\end{CodeInput}
\end{CodeChunk}

For illustrative purposes, the R script of this example includes a \code{gmjmcmc} run with only $P = 3$ populations, which results in the predictor \code{Base} having a posterior probability close to 1. The full analysis is carried out using the parallel version with 40 chains and $P = 25$ populations per chain.
\begin{CodeChunk}
\begin{CodeInput}
R> result2 <- fbms(formula = y ~ 1+., data = df, transforms = transforms,
R>                 probs = probs, params = params, P=25, N = 100,
R>                 method = "gmjmcmc.parallel", runs = 40, cores = 40,
R>                 family = "custom", loglik.pi = poisson.loglik.inla,
R>                 model_prior = list(r = 1/dim(df)[1]), 
R>                 extra_params = list(PID = data$Ind, INLA.num.threads = 1))
R> summary(result2, labels = names(df)[-1], tol = 0.01)
\end{CodeInput}
\end{CodeChunk}

\begin{CodeChunk}
\begin{CodeOutput}
  feats.strings marg.probs
1  p0pm05(Base) 0.86954500
2          Base 0.11206497
3            V4 0.03419890
4     p05(Base) 0.01854366
\end{CodeOutput}
\end{CodeChunk}
This means that the majority of the posterior mass is on the nonlinear transformation $\log(\mbox{Base}) / \sqrt{\mbox{Base}}$ of the baseline number of seizures. Note that for this example the runtime was approximately one hour.


\subsection[sub]{Example 11: Subsampling}\label{SubSec:sub}

In the case of tall data, that is datasets with a large sample size, the computation of maximum likelihood estimates for models explored by the GMJMCMC algorithm can become a computational bottleneck. To speed up this process, \citet{lachmann2022subsampling} introduced a subsampling approach, where only a subset of the available data is used during each iteration of the optimization algorithm. We will demonstrate how this approach can be integrated into the \pkg{FBMS} framework.

As an example, we consider the Heart Disease Health Indicators Dataset, which is available on the Kaggle website
\href{https://www.kaggle.com/datasets/alexteboul/heart-disease-health-indicators-dataset}{www.kaggle.com}.
This dataset contains survey responses from 253,680 individuals, with the primary outcome variable indicating the presence of heart disease (Yes/No). It includes 21 potential predictor variables, such as demographic characteristics (e.g., sex, age, education level), known medical risk factors (e.g., blood pressure, BMI, cholesterol levels), and various lifestyle and dietary habits.

We first load the data and set up the parameters for \code{gmjmcmc} as before, with one additional parameter specifying the percentage of data to be used in the subsampling scheme. In this example, we use only 1\% of the data when optimizing the maximum likelihood (for algorithmic details, see \cite{lachmann2022subsampling}), so we set \code{params$loglik$subs = 0.01}. Additionally, we use a reduced set of fractional polynomials along with the \code{sigmoid} function as potential nonlinear transformations.

\begin{CodeChunk}
\begin{CodeInput}
R> library(RKaggle)
R> df <- RKaggle::get_dataset("alexteboul/heart-disease-health-indicators-dataset")
R>
R> params <- gen.params.gmjmcmc(data = df)
R> params$loglik$subs = 0.01
R> transforms <- c("sigmoid","pm1","p0","p05","p2","p3")
R> probs <- gen.probs.gmjmcmc(transforms)
\end{CodeInput}
\end{CodeChunk}

Next, we specify a new function to compute an approximation of the marginal model posterior. In this function, a subsampling scheme is applied within the iteratively reweighted least squares (IRLS) approximation used to compute the maximum likelihood estimate of the logistic regression model. To implement this, we utilize the \code{irls.sgd} function available on GitHub, which is fully described in \cite{lachmann2022subsampling}.

In brief, we combine a subsampling-based stochastic optimization method for computing marginal likelihoods with MJMCMC. The marginal likelihood estimate is updated each time the model is revisited. This results in a time-inhomogeneous MCMC (TIMCMC) framework with proven convergence properties, assuming that the optimization routine converges. This assumption is satisfied by stochastic gradient descent (SGD) combined with subsampling iteratively reweighted least squares (S-IRLS-SGD). In this optimization routine, the S-IRLS step ensures rapid convergence near local optima, which is further refined by SGD, providing convergence guarantees. Because GMJMCMC is a recurrent algorithm, it directly benefits from this framework, enabling efficient exploration of complex posterior distributions while reducing computational cost.

When implementing the marginal likelihood estimator, we make specific choices of tuning parameters for combining S-IRLS-SGD with GMJMCMC. These include the subsample size (\code{subs}) used in the optimization steps in both the \code{irls.controls} and \code{sgd.controls} lists, as well as the number of S-IRLS and SGD iterations per model visit (\code{maxit}) for both functions. In addition, one must specify the convergence tolerance (\code{tol}) and the cooling schedule for the step size (\code{cooling}) in \code{irls.controls}. For \code{sgd.controls}, the learning rate (\code{alpha}) and its decay parameter (\code{decay}) also need to be specified.

In the function below, we estimate the model \code{mod} using \code{irls.sgd}, then compute the Laplace approximation of the marginal likelihood (\code{mloglik}) and  the logarithm of the model prior (\code{lp}). As in previous examples, the function returns a list containing the marginal log-posterior of the model (\code{crit = mloglik + lp}) and the posterior modes of coefficients (\code{mod$coefficients}).

\begin{CodeChunk}
\begin{CodeInput}
R> devtools::install_github("jonlachmann/irls.sgd", force=T, build_vignettes=F)
R> library(irls.sgd)
R> 
R> logistic.posterior.bic.irlssgd <- function (y, x, model, complex, params) { 
R>   if (!is.null(mlpost_params$crit)) {# First visit of a model
R>      mod <- glm.sgd(x[,model], y, binomial(), 
R>            sgd.ctrl = list(start=mlpost_params$coefs, subs=mlpost_params$subs, 
R>                       maxit=10, alpha=0.00008, decay=0.99, histfreq=10))
R>     mod$deviance <- get_deviance(mod$coefficients, x[,model], y, binomial())
R>     mod$rank <- length(mod$coefficients)
R>  } else {# Warm start at the later visit
R>     mod <- irls.sgd(as.matrix(x[,model]), y, binomial(),
R>              irls.control=list(subs=mlpost_params$subs, maxit=20, 
R>                         tol=1e-7, cooling = c(1,0.9,0.75), expl = c(3,1.5,1)),
R>              sgd.control=list(subs=mlpost_params$subs, maxit=250,
R>                         alpha=0.001, decay=0.99, histfreq=10))
R>  }
R>       # logarithm of marginal likelihood
R>       mloglik <- -mod$deviance /2 - 0.5*log(length(y)) * (mod$rank-1) 
R>       # logarithm of model prior
R>       if (length(params$r) == 0)  params$r <- 1/dim(x)[1]  #default
R>       lp <- log_prior(params, complex)
R>         
R>       return(list(crit = mloglik + lp, coefs = mod$coefficients))
R> }
\end{CodeInput}
\end{CodeChunk}

We begin by comparing the runtime of the subsampling approach with that of regular ML estimation, using a single thread with $P = 2$ populations. In this example, we set $r = 0.01$ for the parameter in the model prior from Equation~(\ref{glmgammaprior}). This choice imposes a weaker penalty on more complex models compared to the default $r=1/n$, which would be overly restrictive given the large number of observations.

\begin{CodeChunk}
\begin{CodeInput}
R> result1 <- fbms(formula = HeartDiseaseorAttack ~ 1 + ., data = df, P = 2,
R>            transforms = transforms, params = params, method = "gmjmcmc",
R>            family = "custom", loglik.pi = logistic.posterior.bic.irlssgd,  
R>            model_prior = list(r = 0.01, subs = 0.01),  sub  = T)                
R>
R> result2 <- fbms(formula = HeartDiseaseorAttack ~ 1 + ., data = df, P = 2,
R>            transforms = transforms, params = params, method = "gmjmcmc", 
R>            family = "binomial", beta_prior = list(type = "Jeffreys-BIC"),
R>            model_prior = list(r = 0.01))
\end{CodeInput}
\end{CodeChunk}

On our cluster, the subsampling approach required a bit more than 2 minutes, whereas regular ML estimation took nearly 8 minutes, which is almost four times longer. It should be noted, however, that the optimization routines in the \pkg{FBMS} package are highly optimized, with the most time-consuming procedures implemented in C++, while the \code{irls.sgd} function for the subsampling approach is primarily implemented in R. With a more efficient implementation of \code{irls.sgd}, the runtime advantage of the subsampling approach could be further increased. The results from these two runs are not reliable, as running a single thread with only two populations means that GMJMCMC has certainly not converged. The purpose of this initial analysis was to demonstrate how to run subsampling and to compare runtimes.

We next consider the results from a slightly more extensive analysis, running 10 parallel chains with $P=3$ population updates. While this setting is still somewhat limited for a dataset of this size, it entails still a managable runtime with about 6 minutes for the subsampling approach and 25 minutes for regular marginal likelihood estimation.

\begin{CodeChunk}
\begin{CodeInput}
R> result_parallel_1 <- fbms(formula = HeartDiseaseorAttack ~ 1 + ., 
R>           data = df, P = 3, transforms = transforms, params = params, 
R>           method = "gmjmcmc.parallel", runs = 10, cores = 10,
R>           family = "custom", loglik.pi = logistic.posterior.bic.irlssgd, 
R>           model_prior = list(r = 0.01, subs = 0.01), sub  = T)
R>
R> result_parallel_2 <- fbms(formula = HeartDiseaseorAttack ~ 1 + .,  
R>           data = df, P = 3, transforms = transforms, params = params, 
R>           method = "gmjmcmc.parallel", runs = 10, cores = 10,
R>           family = "binomial", beta_prior = list(type = "Jeffreys-BIC"),
             model_prior = list(r = 0.01))
\end{CodeInput}
\end{CodeChunk}

The detailed results are presented in Appendix~\ref{SubSec:fbms}. However, these findings remain somewhat unstable, as repeated runs of \code{fbms} with the current settings yield noticeably different outcomes. To address this, we conducted an additional analysis using 40 parallel runs with $P = 10$ populations and $N = 500$ MJMCMC iterations. This more robust configuration increased the runtime to approximately 1.5 hours using subsampling and 4.5 hours with the regular estimation approach. The corresponding detailed results are also provided in Appendix~\ref{SubSec:fbms}.

In this final analysis, the results from the subsampling approach and the regular analysis are not identical but broadly similar. Both methods select 11 linear features, 10 of which overlap. The subsampling approach includes \code{Income}, which is not selected by the regular analysis. Instead, the regular analysis includes the nonlinear feature \code{p3(p05(1+1*Income))} of depth 3, representing a modification of a very simple nonlinear projection of \code{Income},
along with several interaction terms involving \code{Income}.  Conversely, the regular analysis selects \code{CholCheck}, which is not identified by the subsampling approach.

In addition to linear terms, both methods identify several interaction effects, many of which involve \code{Age}. Notably, both approaches include the interaction term \code{Stroke*Age}. Other interaction terms differ between the two methods but show conceptual similarities. For instance, the subsampling approach includes \code{HighChol*Diabetes}, while the regular analysis identifies \code{(Diabetes*HighBP)*HighChol}.


\subsection[Cox]{Example 12: Cox Regression}\label{SubSec:Survival}
As an example of survival data, we use the GSGB breast cancer dataset, which was extensively described by \citet{royston2008multivariable}. The dataset contains survival information for 686 patients, including 299 events, along with 9 covariates serving as potential predictors. Our analysis is based on Cox proportional hazards regression, implemented using the \pkg{survival} package in R.

After downloading the data, we prepare it for analysis by placing the variables representing survival time and censoring status in the first two columns of the data frame \code{df}, and retaining the 9 covariates listed in Table~\ref{Tab:CovEx12} as potential predictors. The original variable $x_{5e}$, which is simply a nonlinear transformation of $x_5$, is removed. We then split the dataset, using two-thirds of the patients as training sample and the remaining third as test sample, ensuring that the proportion of events is preserved in both subsets.
\begin{table}[htb!]
\begin{center}
\caption{List of covariates from the GSGB breast cancer data set as descibed by \citet{royston2008multivariable} in Appendix A.3.}
\label{Tab:CovEx12}
\ \\
\begin{tabular}{lll}
\hline
Variable & Name & Details\\
\hline
$x_1$  & age & Age (years) \\
$x_2$  & meno & Menopausal status (0 = premeno, 1 = postmeno) \\
$x_3$  & size & Tumour size (mm) \\
$x_{4a}$  & gradd1 & 0 = tumour grade 1, $\quad$ \ \ 1 = tumour grade 2 or 3 \\
$x_{4b}$  & gradd2 & 0 = tumour grade 1 or 2,  1 = tumour grade 3 or 3  \\
$x_5$  & nodes & Number of positive lymph nodes \\
$x_6$  & pgr &  Progesterone receptor status (fmol/l) \\ 
$x_7$  & er&  Oestrogen receptor status (fmol/l)\\
$x_8$  & hormon & Tamoxifen treatment (o = no, 1 = yes) \\
\hline
\end{tabular}
\end{center}
\end{table}

To run \code{gmjmcmc}, the input dataframe must contain only a single outcome variable along with the predictor covariates. In survival analysis, however, the outcome is represented by two variables: time to event and censoring status. To address this, we store the \code{time} variable separately, exclude it from the dataframe passed to \code{gmjmcmc}, and instead provide it via \code{extra_params} when calling \code{fbms}. In addition, when generating \code{params}, we must ensure that the correct number of covariates is specified.
\begin{CodeChunk}
\begin{CodeInput}
R> time <- df.train$time
R> params <- gen.params.gmjmcmc(ncol(df.train) - 2)
\end{CodeInput}
\end{CodeChunk}
For \code{transforms}, we use first-order fractional polynomials. We are now ready to define the custom function for approximating the log marginal posterior in the Cox regression model.
\begin{CodeChunk}
\begin{CodeInput}
R> surv.pseudo.loglik = function(y, x, model, complex, mlpost_params)
R> {
R>   data <- data.frame(time = mlpost_params$time, cens = y,
R>                      as.matrix(x[,model]))[,-3]  # Removing intercept
R>
R>   # Fitting Cox model
R>   if(dim(data)[2]==2)    #Take care of the null model
R>   {  
R>      formula1 <- as.formula(paste0("Surv(time,cens)","~ 1"))
R>      out <- coxph(formula1, data = data)
R>      out$loglik <- c(out$loglik,out$loglik)
R>      out$coefficients <- NULL
R>   }  else {
R>      formula1 = as.formula(paste0("Surv(time,cens)","~ 1 + ."))  
R>      out = coxph(formula1, data = data)
R>     
R>      # logarithm of marginal likelihood
R>      mloglik <- (out$loglik[2] - out$loglik[1]) 
R>                   - log(length(y)) * (dim(data)[2] - 2)/2   
R>     
R>      # logarithm of model prior
R>      if (length(mlpost_params$r) == 0) mlpost_params$r <- 1/dim(x)[1] 
R>      lp <- log_prior(mlpost_params, complex)
R> 
R>      # Compute criterion and consider multicollinearity 
R>      crit <- mloglik + lp
R>      if(sum(is.na(out$coefficients))>0)   
R>                       crit <- -.Machine$double.xmax
R>      return(list(crit = crit, coefs =  c(0,out$coefficients)))
R>   }
R> }
\end{CodeInput}
\end{CodeChunk}
The implementation is fairly straightforward, making use of the fact that \code{coxph} provides both the (pseudo) log-likelihood of the current model and that of the baseline model. By default, the \code{fbms} function checks for pairwise linear collinearity (see Appendix~\ref{subset:gen.probs} for a description of \code{params$feat$check.col}). However, in this example, multicollinearity involves more than two features. To prevent such models from being considered, we additionally check whether \code{coxph} returns any coefficients with \code{NA} values.

We compare four different modeling approaches:
\begin{enumerate}
\item[M1:] Single-chain analysis with $P = 5$ and the default settings for nonlinear operators.
\item[M2:] 40 parallel chains including only linear features, using \code{mjmcmc.parallel}.
\item[M3:] 40 parallel chains including only fractional polynomials of depth~1.
\item[M4:] 40 parallel chains with fractional polynomials, allowing also for interactions and projections.
\end{enumerate}

For the purpose of illustration, we specify and run model M4 as follows.
\begin{CodeChunk}
\begin{CodeInput}
R> probs$gen <- c(1,1,1,1)
R> result4 <- fbms(formula = cens ~ 1 + .,data = df.train[,-1],  P = 10, 
R>                 probs = probs, params = params, transforms = transforms,
R>                 method = "gmjmcmc.parallel", runs = 40, cores = 40,
R>                 family = "custom", loglik.pi = surv.pseudo.loglik, 
R>                 model_prior = list(r = 0.5), 
R>                 extra_params = list(time = time))
\end{CodeInput}
\end{CodeChunk}
For the three nonlinear models we set $r = 0.5$ in the model prior, a relatively large value that imposes only mild penalties on nonlinear features. 

The results for the four parameter settings are presented in Appendix~\ref{SubSec:fbms}. Model 1 includes nonlinear modifications of \code{pgr} and \code{nodes} with large posterior probabilities, as well as \code{gradd1} with a posterior probability of 0.19. These findings largely align with those of \citet{royston2008multivariable}, who additionally suggested that a fractional polynomial of \code{age} might be relevant. Model 2, which includes only linear predictors, again identifies \code{pgr} and \code{nodes} with large posterior probabilities, but now both \code{gradd1} and \code{gradd2} appear with intermediate posterior probabilities. The fractional polynomial analysis of Model 3 consistently selects the nonlinear features $\sqrt{pgr}$ and $\log(nodes)$ as having the highest posterior probabilities. Model 4 yieldes similar features, but with some posterior dilution; different runs of \code{fbms} tend to produce varying results. This instability appears to stem from substantial multicollinearity in the data when a wide range of nonlinear features is introduced. Note that if two transformations of the same covariate have similar rank orderings within risk sets, they will provide almost the same contribution to the score equations.

We also want to evaluate the predictive performance on the test dataset of the different models fitted to the training data. To compare the different approaches, we compute the concordance index for right-censored survival times using the \pkg{pec} package. For reference, we include a model with all covariates included as linear predictors and a null model containing only the intercept.

\begin{table}[htb!]
\begin{center}
\caption{C-index values for six models applied to the GSGB breast cancer dataset. Models 1–4 were fitted using \code{fbms} with different feature specifications (see main text), while Full Model includes all covariates as linear predictors and Null Model contains only an intercept. The listed prediction methods (Model Averaging, Best Model, and MPM) apply only to Models 1–4.}
\label{Tab:ResEx12}
\ \\
\begin{tabular}{l|llllll}
\hline
Method & Model 1 & Model 2 & Model 3 & Model 4 & Full Model & Null Model\\
\hline
 Model Averaging  & 0.666  & 0.641  & 0.687 &  0.687   &          0.643  &      0.5\\
Best Model  &      0.662  & 0.624 &  0.686  & 0.683   &          0.643   &     0.5\\
MPM        &       0.662 &  0.648  & 0.675  & 0.593 &            0.643  &      0.5 \\
\hline
\end{tabular}
\end{center}
\end{table}
The results in Table \ref{Tab:ResEx12} present concordance index (C-index) values, a measure of predictive accuracy for right-censored survival data, where 1 indicates perfect prediction and 0.5 corresponds to random guessing (as seen for the Null model). Overall, model averaging consistently yields better predictive performance compared to selecting the single best model or the median probability model (MPM), except for Model 2 where the MPM slightly outperforms the best model.

Model 2, which uses only linear features and was obtained via \code{mjmcmc.parallel}, performs comparably to Full Model (all covariates included linearly), but substantially worse than Models 1, 3, and 4 that incorporate nonlinear features. Models 3 and 4, both based on 40 parallel chains with nonlinear features, achieve the highest model-averaged C-index of 0.687, whereas Model 1, a single-chain analysis, shows slightly lower predictive performance (0.666). While the best models of Models 3 and 4 perform nearly equally well, the MPM results reveal a marked difference: Model 3’s MPM maintains a high C-index of 0.675, whereas Model 4’s MPM drops substantially to 0.593. This decline reflects the posterior dilution observed when allowing more complex nonlinear and interaction terms in Model 4, resulting in greater uncertainty about variable inclusion.

\section[Discussion]{Discussion}\label{Sec:Discussion}

The \pkg{FBMS} framework offers a flexible and extensible approach to Bayesian model selection and model averaging, with three key features justifying the term flexible in its name.
First, as demonstrated in Section~\ref{Sec:Metric}, the methodology supports a wide range of nonlinear feature generation techniques—such as fractional polynomials, logic regression terms, and functional trees—enabling the construction of rich predictor spaces that capture complex data relationships without requiring explicit manual specification.
Second, for generalized linear models (GLMs), \pkg{FBMS} incorporates a broad selection of prior distributions for coefficients, as detailed in Sections~\ref{SubSec:PriorSpec} and~\ref{Sec:Priors}, allowing users to align prior choices with domain expertise, regularization objectives, and interpretability considerations.
Third, the package provides an interface (Section~\ref{Sec:Extensions}) that facilitates extension beyond standard GLMs by permitting the specification of custom likelihoods and prior structures, thereby expanding the framework’s applicability to a wide array of scientific problems.

The twelve examples presented illustrate how to use \pkg{FBMS} to achieve competitive predictive performance while preserving interpretability through posterior inclusion probabilities and model averaging. The package’s modular design supports flexible specification of likelihood functions and priors, enabling adaptation to a wide range of modeling scenarios, from Gaussian and logistic regression to user-defined mixed effects and survival models. This versatility makes \pkg{FBMS} suitable not only for inferential and predictive tasks but also for exploratory analyses where quantifying model uncertainty is essential.

Some challenges remain. Computational demands grow with the size and complexity of the feature space, and model performance is sensitive to algorithmic tuning parameters. While parallelization and subsampling improve scalability, further methodological advances, such as adaptive search strategies, automatic parameter tuning \citep{hubin2019adaptive}, and diagnostics for convergence and exploration efficiency, would enhance usability. It is also planned, that the package will incorporate more efficient inference algorithms, including stochastic evolutionary variational Bayes \citep{sommerfelt2024evolutionary} and BAS \citep{Clyde:Ghosh:Littman:2010}, alongside methods which still need to be newly developed. Further developments may focus on seamless integration of \pkg{FBMS} with distributed computing environments, extended support for longitudinal and streaming data, and coupling nonlinear feature generation with structured prior information in application domains like genomics and epidemiology.

In summary, \pkg{FBMS} achieves flexibility through its ability to generate diverse nonlinear predictors, support a broad range of prior specifications, and interface seamlessly with complex model structures. Together with efficient Bayesian inference algorithms, these features make \pkg{FBMS} a powerful and versatile tool for addressing model uncertainty in complex regression problems. We believe this combination sets it apart as a unique and valuable addition to the statistical software toolbox.

\newpage

\appendix

\section{Appendix: Computational Details} \label{Sec:CompDet}

\subsection{Choice of probabilities:  gen.probs.mjmcmc() and gen.probs.gmjmcmc()
 }\label{subset:gen.probs}

In this section, we provide a detailed description of the key probabilities involved in using \code{mjmcmc} and \code{gmjmcmc}. Both functions include an argument \code{probs}, which is a list of probabilities controlling various algorithmic choices. This list can be created using the helper functions \code{gen.probs.mjmcmc} and \code{gen.probs.gmjmcmc}, which generate appropriate default values. Users can then customize \code{probs} by modifying specific elements of this list. If \code{probs} is not supplied when calling \code{mjmcmc} or \code{gmjmcmc}, the functions will use these default settings.

\subsubsection{gen.probs.mjmcmc} 
This takes no arguments and generates the following list;

\begin{CodeChunk}
\begin{CodeInput}
R> probs <- gen.probs.mjmcmc()
R> str(probs)
\end{CodeInput}
\end{CodeChunk}

\begin{CodeChunk}
\begin{CodeOutput}
$large
[1] 0.05

$large.kern
[1] 0 0 0 1

$localopt.kern
[1] 0.5 0.5

$random.kern
[1] 0.5 0.5

$mh
[1] 0.2 0.2 0.2 0.2 0.1 0.1
\end{CodeOutput}
\end{CodeChunk}

The list consists of five elements, each relating to specific details of the MJMCMC algorithm. 

\begin{itemize}
    \item \code{probs$large}: \ \ The probability of proposing a large jump in the MJMCMC algorithm. With this probability, a large jump proposal is made; otherwise, a local Metropolis-Hastings proposal is used. When specifying this parameter, it is important to balance good mixing around modes and effective exploration between modes.

    \item \code{probs$large.kern}: \ \ A vector of probabilities for different types of large jumps, as described in \citet{hubin2018mode}. The first element corresponds to a random change with a random neighborhood size, the second to a random change with a fixed neighborhood size, the third to a swap with a random neighborhood size, and the fourth to a swap with a fixed neighborhood size. See Table 1 in \citet{hubin2018mode} for further details. If the vector is not normalized to sum to one, it will be normalized automatically.
    
    \item \code{probs$localopt.kern}: \ \ \ A vector of probabilities for the types of local optimizers used during large jumps. The first element corresponds to simulated annealing, and the second to a greedy optimizer. If the two components are not normalized to sum to one, normalization will be done automatically.
    
    \item \code{probs$random.kern}: \ \  A vector of probabilities for the types of randomization kernels applied after local optimization. The four components correspond to the same kernel types as in \code{large.kern}, but are used for local proposals with varying neighborhood sizes. The sizes of these neighborhoods are defined in the \code{params} argument, which we describe later in Appendix \ref{subset:gen.params}.
    
    \item \code{probs$mh}: \ \ A numeric vector specifying the probabilities of different standard Metropolis-Hastings kernels. The first four components correspond to the same kernels as above, while the fifth and sixth components correspond to uniform addition or deletion of a covariate.
    
\end{itemize}

\subsubsection{gen.probs.gmjmcmc} 
This takes the argument \code{transforms} which is a vector with the names of nonlinear functions in the set $\mathcal{G}$. It generates the following list;

\begin{CodeChunk}
\begin{CodeInput}
R> probs <- gen.probs.gmjmcmc(transforms)
R> str(probs)
\end{CodeInput}
\end{CodeChunk}

\begin{CodeChunk}
\begin{CodeOutput}
List of 8
 $ large        : num 0.05
 $ large.kern   : num [1:4] 0 0 0 1
 $ localopt.kern: num [1:2] 0.5 0.5
 $ random.kern  : num [1:4] 0.5 0.5
 $ mh           : num [1:6] 0.2 0.2 0.2 0.2 0.1 0.1
 $ filter       : num 0.6
 $ gen          : num [1:4] 0.25 0.25 0.25 0.25
 $ trans        : num [1:6] 0.167 0.167 0.167 0.167 0.167 ...
\end{CodeOutput}
\end{CodeChunk}

The first five elements of the list concern details of the MJMCMC algorithm and have been discussed above. The last three elements are crucial for the genetic algorithm component of \code{gmjmcmc} and are defined as follows:

\begin{itemize}
    \item \code{probs$filter}: \ \ This controls the removal of features with low posterior probability from the current population of features. Per default, it is set to 0.6. This means that only features with posterior probabilities below 0.6 are considered for removal, with removal probability proportional to one minus their marginal inclusion probability within the population.

    \item \code{probs$gen}: \ \ This is the most important element in the \code{probs} list, as it determines the probabilities of applying the four different operators used to generate new nonlinear features, thereby shaping the feature space. The first entry corresponds to the probability of creating an interaction, followed by modification, nonlinear projection, and finally the mutation operator. The mutation operator reintroduces covariates previously discarded by randomly selecting features from the discarded list. If the four entries of \code{probs$gen} do not sum to one, they are normalized to form a valid probability distribution.

    \item \code{probs$trans}: \ \ The final list element specifies the probabilities with which nonlinear functions are chosen from the set $\mathcal{G}$. By default, a uniform distribution over all functions provided in \code{transforms} is used. To be able to customize these probabilities, \code{transforms} must be passed as an argument to the function.
    
\end{itemize}


\subsection{Choice of further parameters: gen.params.mjmcmc() and gen.params.gmjmcmc()}\label{subset:gen.params}

Apart from the probabilities specified in the \code{probs} list, there are additional hyperparameters and tuning parameters that control the behavior of the \code{mjmcmc} and \code{gmjmcmc} algorithms. These parameters are most conveniently set using the functions \code{gen.params.mjmcmc} and \code{gen.params.gmjmcmc}. We will first examine the former.

\subsubsection{gen.params.mjmcmc} 
This function takes the number of covariates in the dataset as an argument and generates the following list, where some list elements are themselves lists:

\begin{CodeChunk}
\begin{CodeInput}
R> params <- gen.params.mjmcmc(ncol(df)-1)
R> str(params)
\end{CodeInput}
\end{CodeChunk}

\begin{CodeChunk}
\begin{CodeOutput}
 $ burn_in: num 100
 $ mh     :List of 3
  ..$ neigh.size: num 1
  ..$ neigh.min : num 1
  ..$ neigh.max : num 2
 $ large  :List of 3
  ..$ neigh.size: num 35
  ..$ neigh.min : num 25
  ..$ neigh.max : num 45
 $ random :List of 1
  ..$ prob : num 0.01
 $ sa     :List of 5
  ..$ t.init: num 10
  ..$ t.min : num 1e-04
  ..$ dt    : num 3
  ..$ M     : num 12
  ..$ kern  :List of 4
  .. ..$ probs     : num [1:6] 0.1 0.05 0.2 0.3 0.2 0.15
  .. ..$ neigh.size: num 1
  .. ..$ neigh.min : num 1
  .. ..$ neigh.max : num 2
 $ greedy :List of 3
  ..$ steps: num 20
  ..$ tries: num 3
  ..$ kern :List of 4
  .. ..$ probs     : num [1:6] 0.1 0.05 0.2 0.3 0.2 0.15
  .. ..$ neigh.size: num 1
  .. ..$ neigh.min : num 1
  .. ..$ neigh.max : num 2
\end{CodeOutput}
\end{CodeChunk}

The list contains six elements, each corresponding to different parameters of the MJMCMC algorithm. These elements are described as follows:

\begin{itemize}
    \item \code{params$burn_in}: \ \ The burn-in period for the MJMCMC algorithm, set to 100 iterations by default.
    
    \item \code{params$mh}: \ \ A list of parameters for the standard Metropolis-Hastings (MH) kernel:
    \begin{itemize}
        \item \code{neigh.size}: Neighborhood size for MH proposals with fixed proposal size, \\default is 1.
        \item \code{neigh.min}: Minimum neighborhood size for proposals with random size, \\default is 1.
        \item \code{neigh.max}: Maximum neighborhood size for proposals with random size, \\default is 2.    \end{itemize}
    
    \item \code{params$large}: \ \ A list of parameters for the large jump kernel:
    \begin{itemize}
        \item \code{neigh.size}: Neighborhood size for large jump proposals with fixed size, default is the smaller of $0.35 \times p$ and $35$, where $p$ is the number of covariates.
        \item \code{neigh.min}: Minimum neighborhood size for large jumps with random size, default is the smaller of $0.25 \times p$ and $25$.
        \item \code{neigh.max}: Maximum neighborhood size for large jumps with random size, default is the smaller of $0.45 \times p$ and $45$.
    \end{itemize}
    
    \item \code{params$random}: \ \ A list containing one parameter for the randomization kernel:
    \begin{itemize}
        \item \code{prob}: A small probability of changing a component around the mode, default is 0.01.
            \end{itemize}
    
    \item \code{params$sa}: \ \ A list of parameters for the simulated annealing kernel:
    \begin{itemize}
        \item \code{probs}: A numeric vector of length 6 specifying proposal probabilities from \cite{hubin2018mode} used in the simulated annealing algorithm.
        \item \code{neigh.size}: Neighborhood size for simulated annealing proposals, default is 1.
        \item \code{neigh.min}: Minimum neighborhood size, default is 1.
        \item \code{neigh.max}: Maximum neighborhood size, default is 2.
        \item \code{t.init}: Initial temperature for simulated annealing, default is 10.
        \item \code{t.min}: Minimum temperature for simulated annealing, default is 0.0001.
        \item \code{dt}: Temperature decrement factor, default is 3.
        \item \code{M}: Number of iterations in the simulated annealing process, default is 12.
    \end{itemize}
    
    \item \code{params$greedy}: \ \ A list of parameters for the greedy algorithm:
    \begin{itemize}
        \item \code{probs}: A numeric vector of length 6 specifying proposal probabilities from \cite{hubin2018mode} used in the greedy algorithm.
        \item \code{neigh.size}: Neighborhood size for greedy algorithm proposals, default set to 1.
        \item \code{neigh.min}: Minimum neighborhood size for greedy proposals, default  set to 1.
        \item \code{neigh.max}: Maximum neighborhood size for greedy proposals, default  set to 2.
        \item \code{steps}: Number of steps in the greedy algorithm, default  set to 20.
        \item \code{tries}: Number of tries in the greedy algorithm, default  set to 3.
    \end{itemize}

\end{itemize}

\subsubsection{gen.params.gmjmcmc} 
This function takes again the number of covariates in the dataset as an argument. 

\begin{CodeChunk}
\begin{CodeInput}
R> params <- gen.params.gmjmcmc(ncol(df)-1)
\end{CodeInput}
\end{CodeChunk}

It generates a list of 8 elements, where the first six correspond to those from \code{gen.params.gmjmcmc} described above. The remaining two elements are structured as follows:

\begin{CodeChunk}
\begin{CodeOutput}
$ feat         :List of 11
  ..$ D                  : num 5
  ..$ L                  : num 15
  ..$ alpha              : chr "unit"
  ..$ pop.max            : num 13
  ..$ keep.org           : logi FALSE
  ..$ prel.filter        : num 0
  ..$ keep.min           : num 0.8
  ..$ eps                : num 0.05
  ..$ check.col          : logi TRUE
  ..$ col.check.mock.data: logi FALSE
  ..$ max.proj.size      : num 15
 $ rescale.large: logi FALSE
\end{CodeOutput}
\end{CodeChunk}

The most important parameters for the user of \code{gmjmcmc} are contained in the sublist \code{params\$feat}, 
which controls the nonlinear feature selection process:

\begin{itemize}
    \item \code{feat\$D}: \ \ Maximum feature depth, with a default of $D = 5$. This sets the maximum number of 
    recursive feature transformations allowed. In most applications, this is rarely a binding restriction. 
    However, e.g. for fractional polynomials, we tend to set $D = 1$.
    
    \item \code{feat\$L}: \ \  Maximum number of features in a model, defaulting to $L = 15$. If complex models 
    are expected, increasing \code{L} may be necessary.
    
    \item \code{feat\$alpha}: \ \ Strategy for generating the $\alpha$ parameters used in nonlinear projections. The options are:
    \begin{itemize}
        \item \code{"unit"}: Sets all components of $\alpha$ to 1 (default).
        \item \code{"deep"}: Optimizes $\alpha$ across all layers of the feature using a gradient free method.
        
        \item \code{"random"}: Draws $\alpha$ from the prior, enabling a fully Bayesian approach converging as the number of populations increases.
    \end{itemize}
    These methods are discussed further in Section \ref{SubSec:BGNLM}.

    \item \code{feat\$pop.max}: \ \ Maximum size of the feature population explored by MJMCMC. For the initial iteration, the $p$ covariates from the data frame \code{df} form the population. Subsequent iterations use a population size capped at \code{as.integer(1.5 * p)} by default, up to a maximum of 100. Increasing \code{pop.max} can improve exploration, especially for models with many features.

    \item \code{feat\$keep.org}: \ \ Logical flag indicating whether to always keep original covariates in every population. By default, \code{FALSE} allows the algorithm to replace original features with generated ones, but the original covariates then can still reenter through mutation operator of GMJMCMC.

    \item \code{feat\$prel.filter}: \ \ Threshold for pre-filtering covariates before generating the first population. The default (\code{0}) disables pre-filtering. Increasing this value may reduce the number of covariates considered initially.

     \item \code{feat\$prel.select}: \ \ Indices of preliminarily selected covariates. Others can reenter through 
    mutations. The default is \code{NULL}, meaning that all covariates are preselected. For high-dimensional 
    variable selection problems, pre-selection is particularly useful.
    
    \item \code{feat\$keep.min}: \ \ Minimum proportion of features retained during population updates, set by default to \code{0.8}. This prevents drastic changes in the population between iterations, ensuring stability.

    \item \code{feat\$eps}: \ \ Threshold for the inclusion probability of features during feature generation. Features with probabilities below \code{eps = 0.05} are typically not generated, thus controlling exploration aggressiveness.

    \item \code{feat\$check.col}: \ \ Logical flag indicating whether to check for pairwise collinearity among features during generation. The default is \code{TRUE} to prevent linearly dependent features in populations. 

    \item \code{feat\$col.check.mock.data}: \ \ Logical flag indicating whether simulated data is used for pairwise collinearity among features during generation. Otherwise a subsample of real data is used. The default is \code{FALSE}. Setting \code{TRUE} is only recommended if all covariates are continuous. If any binary covariates 
    are present, \code{FALSE} should be used.
    
    \item \code{feat\$max.proj.size}: \ \ Maximum number of previously generated features used to create a new feature via nonlinear projection. The default value is \code{15}, allowing for complex feature construction without excessive computational cost.
\end{itemize}

The parameter \code{rescale.large} (outside the \code{feat} list) is a logical flag. If set to \code{TRUE}, 
large values in the data are rescaled at each population to improve numerical stability. By default, this 
option is disabled.


\subsection{Additional code and results for some examples}\label{SubSec:fbms}

\subsubsection{Robust g-prior for Example 7}
As promised in Section \ref{SubSec:model_prior}, we describe how to analyse the logic regression example using a robust g-prior. Here is the corresponding custom function to compute the log marginal posterior.

\begin{CodeChunk}
\begin{CodeInput}
R> library(BAS) #needed for hyper-geometric functions
R> estimate.logic.tcch = function(y, x, model, complex, mlpost_params)
R> {
R>   # Computation of marginal log likelihood
R>   
R>   suppressWarnings({
R>     mod <- fastglm(as.matrix(x[, model]), y, family = gaussian())
R>   })
R>   p.v <- (mlpost_params$n+1)/(mod$rank+1)
R>   
R>   y_mean <- mean(y)
R>   TSS <- sum((y - y_mean)^2)
R>   RSS <- sum(mod$residuals^2)
R>   R.2 <- 1 - (RSS / TSS)
R>   p <- mod$rank #  this is p_m
R>   
R>   mloglik = (-0.5*p*log(p.v) -0.5*(mlpost_params$n-1)*log(1-(1-1/p.v)*R.2) +
R>       log(beta((mlpost_params$p.a+p)/2,mlpost_params$p.b/2)) -
R>       log(beta(mlpost_params$p.a/2,mlpost_params$p.b/2)) +
R>       log(phi1(mlpost_params$p.b/2,(mlpost_params$n-1)/2,
R>                (mlpost_params$p.a+mlpost_params$p.b+p)/2,
R>                mlpost_params$p.s/2/p.v,R.2/(p.v-(p.v-1)*R.2))) - 
R>       hypergeometric1F1(mlpost_params$p.b/2,
R>       (mlpost_params$p.a+mlpost_params$p.b)/2,mlpost_params$p.s/2/p.v,log = T)) 
R>   if(mloglik ==-Inf||is.na(mloglik )||is.nan(mloglik ))
R>     mloglik  = -10000
R>  
R>   # Computation of log of model prior
R>   
R>   wj <- complex$width
R>   lp <- sum(log(factorial(wj))) - sum(wj*log(mlpost_params$p) + (2*wj-2)*log(2))
R>  
R>   logpost <- mloglik + lp + mlpost_params$n
R>  
R>   if(logpost==-Inf)
R>      logpost = -10000
R>   
R>   return(list(crit = logpost, coefs = mod$coefficients))
R> }
\end{CodeInput}
\end{CodeChunk}

The interested reader is referred to \citet{hubin2018novel} for a detailed mathematical definition of the prior and a discussion of its properties. Under this prior, the GMJMCMC inference is carried out as follows (\code{probs} and \code{params} are set as in the main text when using Jeffreys prior):

\begin{CodeChunk}
\begin{CodeInput}
R>  #Example 7. robust g-prior
R>
R>  result.tcch <- fbms(formula = Y2~1+.,data = df.training, 
R>             probs = probs, params = params,
R>             method = "gmjmcmc", transforms = transforms, N = 500, P = 25,
R>             family = "custom", loglik.pi = estimate.logic.tcch,
R>             model_prior = list(p = p, n = n),
R>             beta_prior =  list(p.a = 1, p.b = 1, p.r = 1.5, p.s = 0, p.k = 1))
\end{CodeInput}
\end{CodeChunk}


\subsubsection{Example 9}

In Section \ref{SubSec:MixedModel}, the function \code{mixed.model.loglik.lme4} was presented. Here, we provide the corresponding code for the other two functions used with \pkg{INLA} and \pkg{RTMB}.

\begin{CodeChunk}
\begin{CodeInput}
R> mixed.model.loglik.inla <- function (y, x, model, complex, mlpost_params) 
R> {
R>   if(sum(model)>1)
R>   {
R>     data1 = data.frame(y, as.matrix(x[,model]), mlpost_params$dr)
R>     formula1 = as.formula(paste0(names(data1)[1],"~",paste0(names(data1)[3:(dim(data1)[2]-1)],
R>                           collapse = "+"),"+ f(mlpost_params.dr,model = \"iid\")"))
R>   } else
R>   {
R>     data1 = data.frame(y, mlpost_params$dr)
R>     formula1 = as.formula(paste0(names(data1)[1],"~","1 +
R>                         f(mlpost_params.dr,model = \"iid\")"))
R>   }
R>   
R>   #to make sure inla is not stuck
R>   inla.setOption(inla.timeout=30)
R>   inla.setOption(num.threads=mlpost_params$INLA.num.threads) 
R>   
R>   mod<-NULL
R>   # error handling for unstable libraries that might crash
R>   tryCatch({
R>     mod <- inla(family = "gaussian",silent = 1L,safe = F, data = data1,formula = formula1)
R>   }, error = function(e) {
R>     
R>     # Handle the error by setting result to NULL
R>     mod <- NULL
R>     # Print a message or log the error if needed
R>     cat("An error occurred:", conditionMessage(e), "\n")
R>   })
R>   
R>   # logarithm of model prior
R>  if (length(mlpost_params$r) == 0)  mlpost_params$r <- 1/dim(x)[1]  # default value or parameter r
R>   lp <- log_prior(mlpost_params, complex)
R>   
R>   if(length(mod)<3||length(mod$mlik[1])==0) {
R>     return(list(crit = -10000 + lp,coefs = rep(0,dim(data1)[2]-2)))
R>   } else {
R>     mloglik <- mod$mlik[1]
R>     return(list(crit = mloglik + lp, coefs = mod$summary.fixed$mode))
R>   }
R> }
\end{CodeInput}
\end{CodeChunk}
We do not want to provide an in-depth commentary on the syntax used to call \pkg{INLA} here. For readers unfamiliar with \pkg{INLA}, comprehensive online documentation is available at \href{https://www.r-inla.org/documentation}{https://www.r-inla.org/documentation}. Instead, we highlight the following key points relevant to the implementation of this function.
\begin{enumerate}
    \item As before, we use the \code{extra_params} list to pass information to the function. In this case, we additionally specify the number of parallel threads for \pkg{INLA} using the argument \code{INLA.num.threads}.
    \item INLA is already relatively slow computationally, and in some cases, it may get stuck during model evaluation. To ensure feasibility, we impose an upper time limit for computing the marginal likelihood of a given model by setting \code{inla.timeout = 30}.
    \item We use \code{tryCatch} to handle situations in which INLA crashes, which occurs occasionally. The model that crash are then excluded from the posterior calculations.
\end{enumerate}
The remainder of the \code{mixed.model.loglik.inla} function should be straightforward to follow, as it closely mirrors the structure of \code{mixed.model.loglik.lme4}.

\begin{CodeChunk}
\begin{CodeInput}
R> mixed.model.loglik.rtmb <- function (y, x, model, complex, mlpost_params) 
R> {
R>   z = model.matrix(y~mlpost_params$dr) #Design matrix for random effect
R>   
R>   msize = sum(model)
R>   #Set up and estimate model
R>   dat = list(y = y, xm = x[,model], z = z)
R>   par = list(logsd_eps = 0,
R>              logsd_dr = 0,
R>              beta = rep(0,msize),
R>              u = rep(0,mlpost_params$nr_dr))
R>   
R>   nll = function(par){
R>     getAll(par,dat)
R>     sd_eps = exp(logsd_eps)
R>     sd_dr = exp(logsd_dr)
R>     
R>     nll = 0
R>     # -log likelihood random effect
R>     nll = nll -  sum(dnorm(u, 0, sd_dr, log = TRUE))
R>     mu = as.vector(as.matrix(xm)%*%beta) + z%*%u
R>     nll <- nll - sum(dnorm(y, mu, sd_eps, log = TRUE))
R>     
R>     return(nll)
R>   }
R>   obj <- MakeADFun(nll , par, random = "u", silent = T )
R>   opt <- nlminb ( obj$par , obj$fn , obj$gr, control = list(iter.max = 10))
R>   
R>   # logarithm of model prior
R>   if (length(mlpost_params$r) == 0)  mlpost_params$r <- 1/dim(x)[1] 

# default value or parameter r
R>   lp <- log_prior(mlpost_params, complex)
R>   
R>   mloglik <- -opt$objective - 0.5*log(dim(x)[1])*msize
R>   return(list(crit = mloglik + lp, coefs = opt$par[-(1:2)]))
R> }
\end{CodeInput}
\end{CodeChunk}

Once again, we do not delve into the details of the \pkg{RTMB} package syntax, which is well documented online at:
\hyperlink{https://cloud.r-project.org/web/packages/RTMB/vignettes/RTMB-introduction.html}{https://cloud.r-project.org/web/packages/RTMB/vignettes/RTMB-\\introduction.html}. We simply note that \pkg{RTMB} requires the number of random intercepts \code{nr_dr} to be explicitly specified. This value is passed via the \code{extra_params} list.


\subsubsection{Example 10}

Here is the specification of \code{poisson.loglik.inla} used in Section \ref{SubSec:MixedPoisson}. It is very similar to \code{mixed.model.loglik.inla}. When using \code{gmjmcmc.parallel} it is better to allow INLA to use only one thread for its own computations.

\begin{CodeChunk}
\begin{CodeInput}
R> poisson.loglik.inla <- function (y, x, model, complex, mlpost_params) 
R> {
R>   if(sum(model)>1)
R>   {
R>     data1 <- data.frame(y, as.matrix(x[,model]), mlpost_params$PID)
R>     formula1 <- as.formula(paste0(names(data1)[1],"~",
R>                    paste0(names(data1)[3:(dim(data1)[2]-1)],collapse = "+"),
R>                    "+ f(mlpost_params.PID,model = \"iid\")"))
R>   } else
R>   {
R>     data1 <- data.frame(y, mlpost_params$PID)
R>     formula1 <- as.formula(paste0(names(data1)[1],"~","1 + 
R>                           f(mlpost_params.PID,model = \"iid\")"))
R>   }
R>   
R>   #to make sure inla is not stuck
R>   inla.setOption(inla.timeout=30)
R>   inla.setOption(num.threads=mlpost_params$INLA.num.threads) 
R>   
R>   mod<-NULL
R>   
R>   #error handling for unstable libraries that might crash
R>   tryCatch({
R>     mod <- inla(family = "poisson",silent = 1L,safe = F,
R>                 data = data1,formula = formula1)
R>   }, error = function(e) {
R>     # Handle the error by setting result to NULL
R>     mod <- NULL
R>     # Print a message or log the error if needed
R>     cat("An error occurred:", conditionMessage(e), "\n")
R>   })
R>   
R>   # logarithm of model prior
R>   if (length(mlpost_params$r) == 0)  mlpost_params$r <- 1/dim(x)[1] 
R>   lp <- log_prior(mlpost_params, complex)
R>   
R>   if(length(mod)<3||length(mod$mlik[1])==0) {
R>    return(list(crit = -10000 + lp,coefs = rep(0,dim(data1)[2]-2)))
R>   } else {
R>     mloglik <- mod$mlik[1]
R>     return(list(crit = mloglik + lp, coefs = mod$summary.fixed$mode))
R>   }
R> }
\end{CodeInput}
\end{CodeChunk}


\subsubsection{Results from Example 11}

\textbf{$10$ Parallel runs, $P = 3, N = 100$, with subsampling}
\begin{CodeChunk}
\begin{CodeOutput}
Best   population: 2  thread: 8  log marginal posterior: -60525.07 

             feats.strings marg.probs
1                   HighBP          1
2                CholCheck          1
3                   Stroke          1
4                 Diabetes          1
5                   Fruits          1
6                  GenHlth          1
7                 DiffWalk          1
8                      Age          1
9                  Veggies          1
10             p0(GenHlth)          1
11                HighChol          1
12                     BMI          1
13                  Smoker          1
14            PhysActivity          1
15       HvyAlcoholConsump          1
16             NoDocbcCost          1
17                     Sex          1
18                  Income          1
19              p2(Income)          1
20      sigmoid(Education)          1
21       (HighBP*PhysHlth)          1
22           (Age*GenHlth)          1
23 (AnyHealthcare*GenHlth)          1
24      (HighChol*Veggies)          1
25  ((Age*GenHlth)*Stroke)          1
26            (Age*Income)          1
27       (DiffWalk*Fruits)          1
28          (Sex*DiffWalk)          1
\end{CodeOutput}
\end{CodeChunk}

\textbf{$10$ Parallel runs, $P = 3, N = 100$, regular analysis without subsampling}
\begin{CodeChunk}
\begin{CodeOutput}
Best   population: 3  thread: 7  log marginal posterior: -60421.18 

                   feats.strings marg.probs
1                         HighBP          1
2                       HighChol          1
3                      CholCheck          1
4                         Smoker          1
5                         Stroke          1
6                       Diabetes          1
7                        GenHlth          1
8                       DiffWalk          1
9                            Sex          1
10                           Age          1
11                        Income          1
12                  (HighBP*Age)          1
13                  p2(PhysHlth)          1
14             HvyAlcoholConsump          1
15                   NoDocbcCost          1
16             (HighChol*Income)          1
17              (Smoker*GenHlth)          1
18 ((HighChol*Income)*(Sex*BMI))          1
19         pm1(1+1*Age+1*HighBP)          1
20                  (Stroke*Age)          1
21           p05((Age*DiffWalk))          1
22       p3((PhysHlth*PhysHlth))          1
23                  sigmoid(Age)          1
24     (Sex*(HighChol*Diabetes))          1
25           (HighChol*Diabetes)          1
\end{CodeOutput}
\end{CodeChunk}

\textbf{$40$ Parallel runs, $P = 10, N = 500$, with subsampling}
\begin{CodeChunk}
\begin{CodeOutput}
Best   population: 8  thread: 26  log marginal posterior: -60376.8 

                       feats.strings   marg.probs
1                             HighBP 1.0000000000
2                           HighChol 1.0000000000
3                             Smoker 1.0000000000
4                             Stroke 1.0000000000
5                           Diabetes 1.0000000000
6                            GenHlth 1.0000000000
7                           DiffWalk 1.0000000000
8                                Age 1.0000000000
9                             Income 1.0000000000
10                    (Age*DiffWalk) 1.0000000000
11                 HvyAlcoholConsump 1.0000000000
12                      (Stroke*Age) 1.0000000000
13                (Age*(Sex*Income)) 1.0000000000
14 p05(((CholCheck*Income)*GenHlth)) 1.0000000000
15               (HighChol*Diabetes) 1.0000000000
16                       NoDocbcCost 1.0000000000
17             (Age*(Sex*Education)) 0.0001717437
\end{CodeOutput}
\end{CodeChunk}

\textbf{$40$ Parallel runs, $P = 10, N = 500$, regular analysis without subsampling}
\begin{CodeChunk}
\begin{CodeOutput}
Best   population: 9  thread: 17  log marginal posterior: -60259.13 

                  feats.strings marg.probs
1                        HighBP          1
2                     CholCheck          1
3                        Smoker          1
4                        Stroke          1
5                      Diabetes          1
6             HvyAlcoholConsump          1
7                   NoDocbcCost          1
8                       GenHlth          1
9                      DiffWalk          1
10                          Age          1
11                     HighChol          1
12                 (Age*HighBP)          1
13            (HighChol*Income)          1
14                 (Age*Stroke)          1
15 ((Diabetes*HighBP)*HighChol)          1
16   ((Income*HighChol)*Stroke)          1
17  ((Age*Sex)*p05(1+1*Income))          1
18        (Age*p05(1+1*Income))          1
19      p2((GenHlth*(Age*Sex)))          1
20          p3(p05(1+1*Income))          1
21    (Income*(Income*GenHlth))          1
\end{CodeOutput}
\end{CodeChunk}


\subsubsection{Results from Example 12}

\begin{CodeChunk}
\begin{CodeInput}
> summary(result1, tol = 0.1)
\end{CodeInput}
\end{CodeChunk}
\begin{CodeChunk}
\begin{CodeInput}
Best   population: 2  log marginal posterior: 25.36382 

  feats.strings marg.probs
1      p05(pgr)  1.0000000
2    p05(nodes)  0.9994586
3        gradd1  0.1946058
\end{CodeInput}
\end{CodeChunk}

\begin{CodeChunk}
\begin{CodeInput}
> summary(result2, tol = 0.1)
\end{CodeInput}
\end{CodeChunk}
\begin{CodeChunk}
\begin{CodeInput}
Best log marginal posterior:  17.14489 

  feats.strings marg.probs
1         nodes  0.9996115
2           pgr  0.9888463
3        gradd2  0.4853617
4        gradd1  0.3342886
5        hormon  0.1406956
\end{CodeInput}
\end{CodeChunk}

\begin{CodeChunk}
\begin{CodeInput}
> summary(result3, tol = 0.1)
\end{CodeInput}
\end{CodeChunk}
\begin{CodeChunk}
\begin{CodeInput}
Best   population: 7  thread: 16  log marginal posterior: 26.41905 

  feats.strings marg.probs
1      p05(pgr)  0.9865323
2     p0(nodes)  0.7038477
3           age  0.1634581
4    p05(nodes)  0.1536846
5      pm2(age)  0.1307798
\end{CodeInput}
\end{CodeChunk}

\begin{CodeChunk}
\begin{CodeInput}
> summary(result4, tol = 0.1)
\end{CodeInput}
\end{CodeChunk}
\begin{CodeChunk}
\begin{CodeInput}
Best   population: 4  thread: 2  log marginal posterior: 26.62292 

         feats.strings marg.probs
1             p05(pgr)  0.5586264
2            p0(nodes)  0.4021045
3           p05(nodes)  0.1857823
4 p0(1+1*pgr+1*gradd1)  0.1646611
5        p0(1+1*nodes)  0.1151796
\end{CodeInput}
\end{CodeChunk}

\section*{Acknowledgments}

We would like to thank Dr. Mariana Nold for constructive comments on improving usability of the package. 

\bibliography{refs}

\end{document}